%% file: pp_ppHpHm_PRD_v3.tex
\documentclass[prd,preprint,tightenlines,floatfix,showpacs,
preprintnumbers,nofootinbib,eqsecnum]{revtex4-1}

\usepackage[T1]{fontenc}		

\usepackage{amsmath,amsfonts,amssymb,amstext,mathrsfs}
\usepackage{mathpazo}

\usepackage[dvips]{graphicx}
\usepackage{epsf,float}
\usepackage{revsymb}

\usepackage{dcolumn}
\usepackage{braket}
\usepackage{color,xcolor}
\usepackage{graphicx}
\usepackage{subfigure}
\usepackage{multirow}
\usepackage{tabularx}
\usepackage{pstricks}
\usepackage[section]{placeins}
\usepackage{booktabs}
\usepackage{array}

\usepackage{hyperref}

\newcommand{\bp}{\mbox{\boldmath $p$}}
\newcommand{\bk}{\mbox{\boldmath $k$}}

\begin{document}

\title{Exclusive production of heavy charged Higgs boson pairs\\
in the $p p \to p p H^+ H^-$ reaction at the LHC and a future circular collider}

\author{Piotr Lebiedowicz}
\email{Piotr.Lebiedowicz@ifj.edu.pl}
\author{Antoni Szczurek \footnote{Also at University of Rzesz\'ow, PL-35-959 Rzesz\'ow, Poland.}}
\email{Antoni.Szczurek@ifj.edu.pl}
\affiliation{Institute of Nuclear Physics PAN, PL-31-342 Krak\'ow, Poland}

\vspace{2cm}

\begin{abstract}
We calculate differential cross sections for exclusive production
of heavy charged scalar, weakly interacting particles 
(charged Higgs bosons, charged technipions, etc.) 
via photon-photon exchanges in the $p p \to p p H^+ H^-$ reaction 
with exact $2 \to 4$ kinematics.
We present distributions in rapidities, transverse momenta, and
correlations in azimuthal angles between the protons and between 
the charged Higgs bosons.
As an example, the integrated cross section for $\sqrt{s}$ = 14~TeV
(LHC) is about 0.1~fb 
and about 0.9~fb at the Future Circular Collider (FCC) for $\sqrt{s}$ = 100~TeV
when assuming $m_{H^{\pm}} = 150$~GeV. 
The results are compared with results obtained within
standard equivalent-photon approximation known from the literature. 
We discuss the role of the Dirac and Pauli electromagnetic form factors 
of the proton. 
We have also performed first calculations of cross sections for 
the exclusive diffractive Khoze-Martin-Ryskin mechanism. 
We have estimated limits on the $g_{h H^+ H^-}$ coupling constant within
two-Higgs dublet model based on recent experimental data from the LHC.
The diffractive contribution is, however, much smaller than 
the $\gamma \gamma$ one.
The $Z \gamma$, $\gamma Z$, and $ZZ$ exchanges give even smaller contributions.
Absorption corrections are calculated for the first time
differentially for various distributions.
In general, they lead to a damping of the cross section. 
The damping depends on the $M_{H^{+}H^{-}}$ invariant mass
and on $t$ four-momentum transfers squared.
In contrast to diffractive processes, the larger the collision energy, 
the smaller the effect of absorption.
We discuss a possibility to measure the exclusive production of 
two charged Higgs bosons with the help of so-called ``forward proton 
detectors'' at the LHC experiments.
\end{abstract}


\maketitle

\section{Introduction}

There are several reasons why exclusive reactions are interesting \cite{Lebiedowicz:thesis,Albrow:2010yb}.
One of them is the possibility to search for effects beyond the Standard Model (SM).
The main advantage of exclusive reactions is that background contributions are
strongly reduced compared to inclusive processes.
A good example are searches for exclusive production
of supersymmetric Higgs boson \cite{Heinemeyer:2007tu,Tasevsky:2013iea,Tasevsky:2014cpa}, anomalous boson couplings for 
$\gamma \gamma \to W^+ W^-$ \cite{Pierzchala:2008xc,Schul:2008sr,Kepka:2008yx,Chapon:2009hh} 
or for $\gamma \gamma \to \gamma \gamma$ \cite{Fichet:2013gsa,Fichet:2014uka}.
So far these processes are usually studied in the so-called 
equivalent-photon approximation (EPA) 
(for a description of the method, see e.g.~\cite{Budnev:1974de}).
Within the Standard Model the cross section for the $p p \to p p W^+ W^-$
reaction is about 100~fb at $\sqrt{s} = 14$~TeV \cite{Lebiedowicz:2012gg}. 
Gluon-induced processes could also contribute to the exclusive
production of $W^+ W^-$ \cite{Lebiedowicz:2012gg} 
and $W^{\pm} H^{\mp}$ \cite{Enberg:2011qh} via quark loops.
\footnote{An attractive channel is the associated production 
of a charged Higgs boson with a $W^{\pm}$ boson via $\gamma \gamma$ fusion. 
Since there is no $H^{\pm}W^{\mp}\gamma$ couplings
the $\gamma \gamma \to H^{\pm} W^{\mp}$ associated production process
have no tree-level contribution in the 2HDM and in the MSSM
and occurs only at one-loop level in the lowest order.}
The corresponding cross sections are rather small
mainly due to suppression of Sudakov form factors and the gap survival factor.
The exclusive reactions could be also used in searches for neutral 
technipion in the diphoton final state \cite{Lebiedowicz:2013fta} 
or dilaton \cite{Goncalves:2015oua}.
Here a precise prediction of the cross sections is not possible
as the model parameters are still unknown.

Discovery of the heavy Higgs bosons of
the Minimal Supersymmetric Standard Model (MSSM) 
\cite{Gunion:1989we,Gunion:1992hs,Djouadi:2005gj}
or more generic Two-Higgs Doublet Models (2HDMs) 
(see e.g. \cite{Diaz:2002tp,Branco:2011iw})
poses a special challenge at future colliders.
One of the international projects currently under consideration
is the Future Circular Collider (FCC) \cite{FCC}.
The Higgs sector in both the MSSM and 2HDM contains five states:
three neutral [two $CP$-even ($h$, $H$) and one $CP$-odd ($A$)]
and two charged ($H^{+}$, $H^{-}$) Higgs bosons.
In general, either $h$ or $H$ could correspond to the SM Higgs.
The charged Higgs boson pair production in 
the $\gamma \gamma \to H^{+} H^{-}$ mode was considered in 
\cite{BowserChao:1993ji,Drees:1994zx,Moretti:2002sn}.
In general, the higher-order corrections to 
the $\gamma \gamma \to H^{+} H^{-}$ subprocess decrease 
the tree-level total cross section by about a few percent; 
see \cite{Zhu:1997es,Lei:2005kr}.
Also the associated production $\gamma \gamma \to H^{\pm} W^{\mp}$ 
was discussed in the literature \cite{Zhou:2001wp}.
For a more extensive discussion of charged Higgs boson 
production at the LHC and ILC, see \cite{Kanemura:2014dea}.

There are also extensive phenomenological studies on charged Higgs boson(s) 
production at the LHC in the inclusive reactions via the partonic processes
\cite{Alves:2005kr,Heinemeyer:2013tqa}.
If $m_{H^{\pm}} < m_{t} - m_{b}$, the charged Higgs boson
can be produced in $t \to b H^{+}$ and $\bar{t} \to \bar{b} H^{-}$ decays
from the parent production channel $pp \to t \bar{t}$,
which would compete with the SM process $t \bar{t} \to b W^{+} \bar{b} W^{-}$.
The dominant decay channels in this mass range are 
$H^{\pm} \to \tau \nu_{\tau}$ and $H^{\pm} \to c \bar{s} (\bar{c} s)$.
In the case of a heavy charged Higgs with $m_{H^{\pm}} > m_{t} - m_{b}$, 
there are three major mechanisms:
\begin{itemize}
\item[(a)] 
Associated production with a top quark via the partonic processes
$q \bar{q}, g g \to t b H^{\pm}$
\cite{Belyaev:2001qm,Belyaev:2002eq,Alwall:2004xw,
Peng:2006wv,Nhung:2012er,Cao:2013ud,Flechl:2014wfa}
as well as through the gluon-bottom fusion 
$gb \to t H^{\pm}$ \cite{Gunion:1986pe,Zhu:2001nt,Gao:2002is,Plehn:2002vy,
Berger:2003sm,Weydert:2009vr,Yang:2011jk}.
The sequential decay $H^{+} \to t \bar{b}$ is known as a preferred channel.
But signals in these processes appear together with large QCD backgrounds.
The $H^{\pm} \to W^{\pm} H / W^{\pm} A \to W^{\pm} b \bar{b}$ channels
were analyzed in \cite{Assamagan:2002ne,Coleppa:2014cca}. 
In the latter paper, the $W^{\pm} \tau \bar{\tau}$ decay channel was also considered.
Recently, the $H^{\pm} \to W^{\pm} (H_{obs} \to b \bar{b})$ decay channel
for a SM-like Higgs was studied in \cite{Basso:2012st,Enberg:2014pua}.
This decay channel can be particularly important 
when charged Higgs is produced through the $pp \to t H^{\pm}$ processes.
\item[(b)] 
Associated production with a $W^{\pm}$ boson 
through the $q \bar{q}, g g \to H^{\pm} W^{\mp}$ subprocesses
\cite{Dicus:1989vf,BarrientosBendezu:1998gd,
Moretti:1998xq,BarrientosBendezu:1999vd,
BarrientosBendezu:2000tu,Brein:2000cv,Hollik:2001hy,Asakawa:2005nx,
Eriksson:2006yt,Bao:2011sy,Liu:2013oen}
and associated production of a charged Higgs boson with a $CP$-odd Higgs boson,
i.e. $q \bar{q} \to H^{\pm} A$, was studied in \cite{Kanemura:2001hz,Cao:2003tr}.
\item[(c)] 
Charged Higgs boson pair production via $q \bar{q}, g g \to H^+ H^-$ \cite{Krause:1997rc,BarrientosBendezu:1999gp,Brein:1999sy,Hespel:2014sla}, 
$b \bar{b} \to H^{+} H^{-}$ \cite{HongSheng:2005uy} subprocesses
or in association with bottom quark pairs $q \bar{q}, gg \to b\bar{b} H^{+} H^{-}$ \cite{Moretti:2001pp,Moretti:2003px}.
For more recent studies, see \cite{Aoki:2011wd,Liu:2015mza}.
\end{itemize}

The cross sections for the inclusive reactions strongly depend on the model parameters, 
such as $\tan\beta \equiv v_2/v_1$, the ratio of 
the vacuum expectation values of the two Higgs doublets, and others.
A program on how to limit the relevant parameters, 
based on the collider searches and data from $B$ factories,
was presented, e.g., in \cite{Cornell:2009gg}.  
Another important ingredient of the model is the mass of the charged Higgs boson.
In the MSSM the relation between the masses of the charged Higgs boson 
and $CP$-odd Higgs boson in lowest order is given by
$m_{H^{\pm}}^2 = m_A^2 + m_{W^{\pm}}^2$ 
\footnote{This is particular to the MSSM in lowest order
(is modified by one-loop radiative corrections \cite{Diaz:1991ki}) 
and does not hold in 2HDMs 
or in, e.g., the Next-to-Minimal Supersymmetric Standard Model (NMSSM).}
(for reviews and details, see, e.g.~\cite{Gunion:1989we,Djouadi:2005gj}).

Several experimental searches already placed limitations on the mass
of the charged Higgs bosons.
There is a direct limit of $m_{H^{\pm}} > 78.6$~GeV from the LEP searches
\cite{Searches:2001ac} by its decays $H^{\pm} \to \tau \nu_{\tau}$
and $H^{\pm} \to c \bar{s} (\bar{c} s)$. 
At hadron colliders, the search procedures for a charged Higgs
boson differ in term of its mass range.  
At the Tevatron the searches were mainly focused on the low mass range
$m_{H^{\pm}} < m_{t}$ which can put a constraint to the 2HDM (as an example)
on the small and large $\tan \beta$ regions for 
a charged Higgs boson mass up to $∼160$~GeV \cite{Abazov:2009aa}.
Recent searches at the LHC \cite{Chatrchyan:2012vca,Aad:2012tj,Aad:2012rjx,
Aad:2013hla,Aad:2014kga,CMS:2014pea,CMS:2014cdp}
provide new limitations on the model parameters.
However, still a possible span of parameters is rather large.
For example, in the latest searches ATLAS and CMS put limits 
on the product of branching fractions
$BR(t \to H^{+} b) \times BR(H^{+} \to \tau \nu_{\tau}$),
but there are no model-independent limits on the $H^{\pm}$ mass.
The observed limits are reinterpreted in some MSSM scenarios,
with mass limits around $140-160$~GeV that depend somewhat on $\tan \beta$.
But in other models such as type-I 2HDM, the limit may be weaker.

Other experimental bounds on the charged Higgs mass come from processes 
where the charged Higgs boson enters as a virtual particle,
i.e. participates in loop diagrams.
It is well known that in the type-II 2HDM, 
where the up- and down-type quarks and leptons couple to different doublets, 
the $b \to s \gamma$ transitions 
imposes a strong constraint on the Higgs boson mass $m_{H^{\pm}} \gtrsim 300$~GeV. 
In the type-I 2HDM, instead, all fermions couple to the same doublet and
there is no such strong $b$-physics constraint
(the MSSM is also less sensitive to radiative corrections).
The flavor constraints on the Higgs sector are, however, typically model dependent.
A detailed analysis of precision and flavor bounds 
in the 2HDM can be found, e.g., in \cite{Coleppa:2013dya}. 


\begin{figure}
\includegraphics[width=0.28\textwidth]{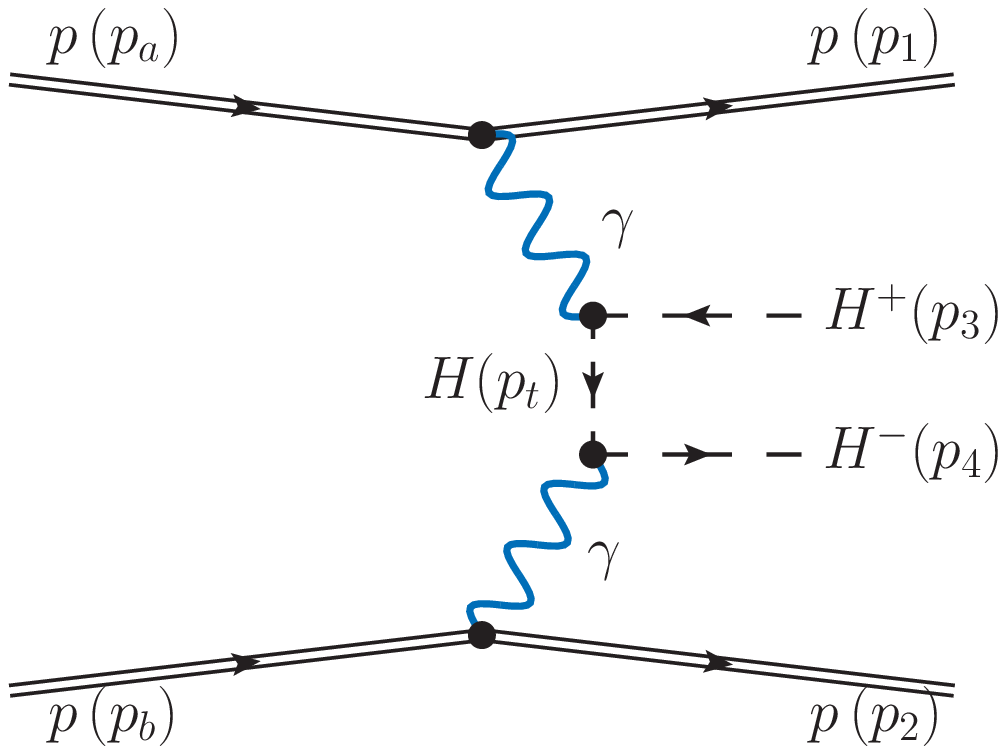}
\includegraphics[width=0.28\textwidth]{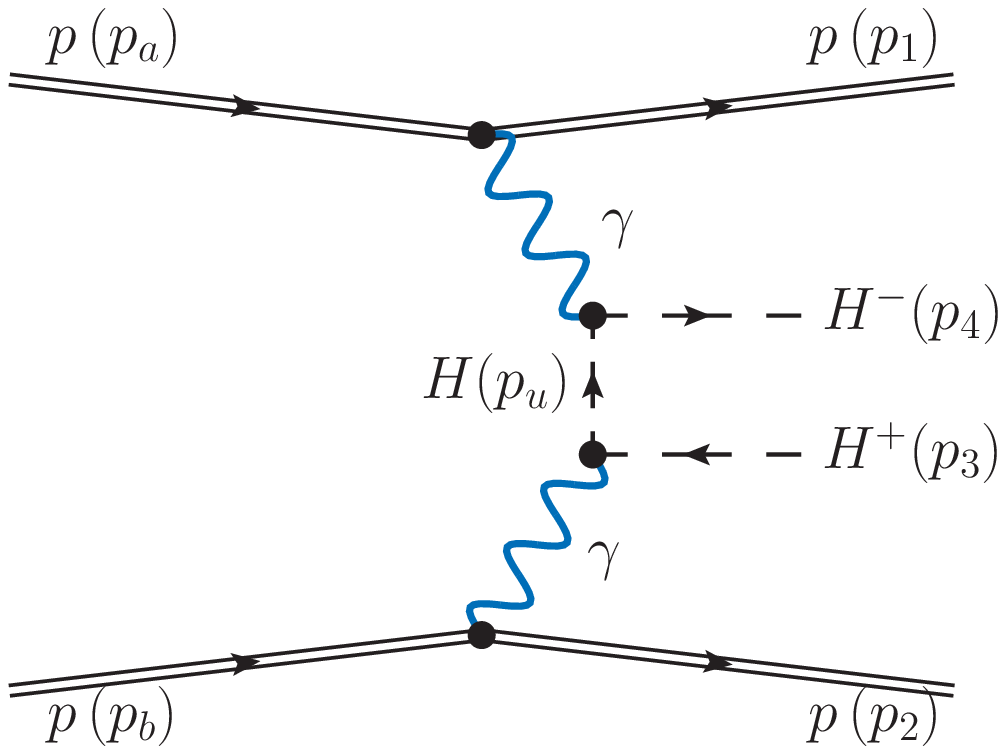}
\includegraphics[width=0.28\textwidth]{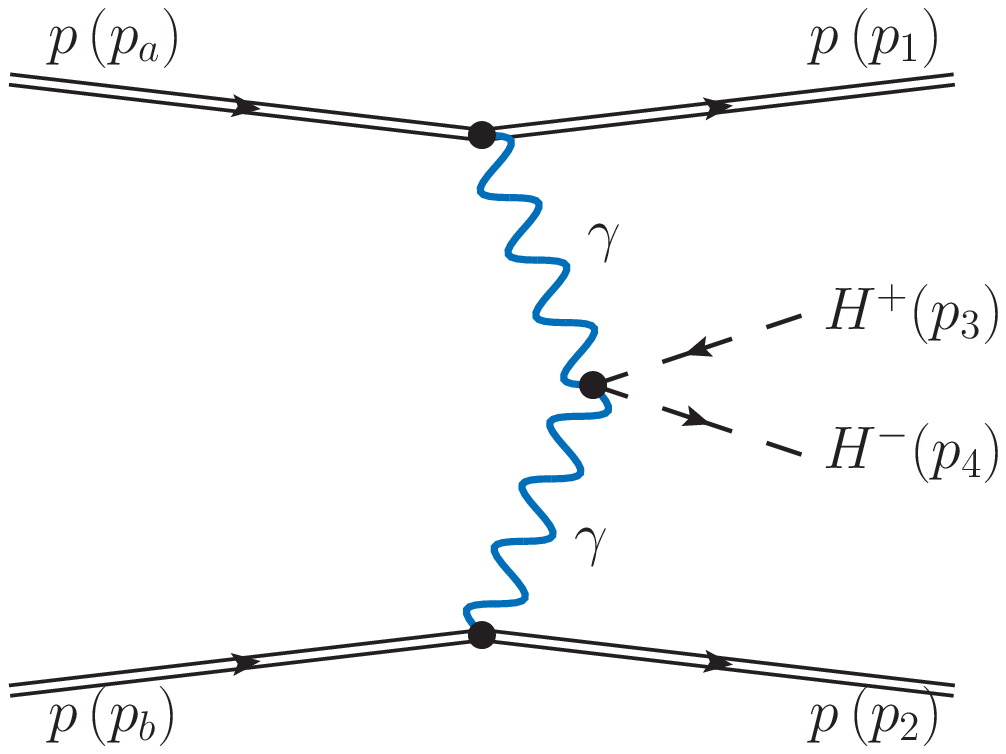}
\caption{Born diagrams for exclusive production of pairs of charged scalar
particles via photon-photon exchanges.}
\label{fig:Born_diagrams}
\end{figure}


In the present analysis we wish to concentrate 
on exclusive production of charged Higgs bosons 
in proton-proton collisions proceeding through exchange of two photons.
In Fig.~\ref{fig:Born_diagrams} we show basic diagrams 
contributing to the $p p \to p p H^+ H^-$ reaction.
The coupling of photons to protons is usually parametrized
with the help of proton electromagnetic form factors: $G_E$ (electric), 
$G_M$ (magnetic) or equivalently $F_1$ (Dirac), $F_2$ (Pauli).
We wish to discuss the dependence on the form factors of several 
differential distributions.
In contrast to inclusive processes discussed above, the considered 
here exclusive reaction is free of the model parameter uncertainties,
at least in the leading order, except of the mass of the charged Higgs bosons.

Our paper is organized as follows.
In Sec.~\ref{sec:section_2} we discuss formalism of the $pp \to pp H^{+} H^{-}$ reaction
both in the equivalent-photon approximation (EPA)
in the momentum space commonly used in the literature
and in exact $2 \to 4$ kinematics.
In Sec.~\ref{sec:section_3} we present numerical results 
for total and differential cross sections.
In Sec.~\ref{sec:electromagnetic_process} we compare results obtained in the 
exact $2 \to 4$ calculation and those obtained in EPA. 
We present not only estimation of the total cross section 
but also several differential distributions important 
for planning potential future experimental searches.
In addition, we discuss the role of absorption corrections 
commonly neglected for two-photon initiated processes.
Finally, we also consider diffractive exclusive production
of the $H^{+} H^{-}$ bosons through an intermediate recently discovered Higgs boson.
Diffractive contribution is discussed in Sec.~\ref{sec:diffractive_process}.

\section{Formalism}
\label{sec:section_2}

We shall study exclusive production of $H^+ H^-$ in proton-proton 
collisions at high energies
\begin{eqnarray}
p(p_{a},\lambda_{a}) + p(p_{b},\lambda_{b}) \to
p(p_{1},\lambda_{1}) + H^{+}(p_{3}) + H^{-}(p_{4}) + p(p_{2},\lambda_{2}) \,,
\label{2to4_reaction}
\end{eqnarray}
where $p_{a,b}$, $p_{1,2}$ and $\lambda_{a,b}$, 
$\lambda_{1,2} = \pm \frac{1}{2}$ 
denote the four-momenta and helicities of the protons, 
and $p_{3,4}$ denote the four-momenta of the charged Higgs bosons, 
respectively.
In the following we will calculate the contributions from the diagrams
of Fig.~\ref{fig:Born_diagrams}.

\subsection{Equivalent-photon approximation}

Similar processes are treated usually in the equivalent-photon
approximation (EPA) in the momentum space, see e.g.
\cite{Lebiedowicz:2012gg,Lebiedowicz:2013fta}.
\footnote{An impact parameter EPA was considered recently in \cite{Dyndal:2014yea}.}
Only very few differential distributions can be obtained in the EPA approach.
In this approximation, when neglecting photon transverse momenta, 
one can write the differential cross section as
\begin{equation}
\frac{d \sigma}{d y_3 d y_4 d^2p_{tH}} = \frac{1}{16 \pi^2 {\hat s}^2} 
x_1  f(x_1) x_2 f(x_2) \overline{|{\cal M}_{\gamma \gamma \to H^+ H^-}|^2} \,,
\label{EPA_formula}
\end{equation}
where ${\hat s} = s x_{1} x_{2}$ and $f(x)$'s are an elastic fluxes of the equivalent photons 
(see e.g.\cite{Budnev:1974de}) 
as a function of longitudinal momentum fraction with respect to the parent proton 
defined by the kinematical variables of the charged Higgs bosons,
\begin{eqnarray}
x_1 = \frac{m_{tH}}{\sqrt{s}}(e^{y_3} + e^{y_4}) \,, \qquad
x_2 = \frac{m_{tH}}{\sqrt{s}}(e^{-y_3} + e^{-y_4}) \,, \qquad 
m_{tH} = \sqrt{|\vec{p}_{tH}|^{2}+m_{H}^{2}}
\label{fractions}
\end{eqnarray}
with $m_{tH}$ being transverse mass of the $H^{\pm}$ boson(s).
Above $\overline{|{\cal M}|^2}$
is the $\gamma \gamma \to  H^+ H^-$ amplitude squared averaged over the photon polarization states.

The photon flux $f(x)$ is given by the formula \cite{Budnev:1974de}
\begin{eqnarray}
f(x) = \dfrac{1}{x} \int_{Q^{2}_{min}}^{Q^{2}_{max}} \frac{\alpha_{em}}{\pi}\frac{dQ^{2}}{Q^{2}}
\left[ \left(1-x\right) \left( 1 - \frac{Q^{2}_{min}}{Q^{2}}\right) D(Q^{2}) + \frac{x^{2}}{2} C(Q^{2})\right]\,,
\label{flux}
\end{eqnarray}
where the spacelike momentum transfer squared 
$Q^{2} \equiv -q^{2} = -t \geqslant 0$ 
\footnote{
Here we discuss the collinear EPA approach,
that is, the photon transverse momenta $\vec{q}_{t} = \vec{0}$.
An approach including transverse momenta of photons was discussed 
recently in \cite{daSilveira:2014jla}.
}
and the photon minimal virtuality allowed by kinematics 
$Q^{2}_{min} = x^{2} m_{p}^{2} (1-x)^{-1}$.
The coefficient functions $C$ and $D$ are
determined by the electric and magnetic form factors of the proton:
\begin{eqnarray}
C(Q^{2}) \equiv G_{M}^{2}(Q^{2})\,, \qquad
D(Q^{2}) \equiv \left( 4 m_{p}^{2} G_{E}^{2}(Q^{2}) + Q^{2} G_{M}^{2}(Q^{2}) \right) 
                \left( 4 m_{p}^{2} + Q^{2}\right)^{-1} \,,
\label{flux_aux1}
\end{eqnarray}
where the $G_{E}$ and $G_{M}$ form factors are related to
Dirac ($F_{1}$) and Pauli ($F_{2}$) form factors by
\begin{eqnarray}
G_{E}(Q^{2}) \equiv F_{1}(Q^{2}) - \frac{Q^{2}}{4 m_{p}^{2}} F_{2}(Q^{2})\,, \qquad
G_{M}(Q^{2}) \equiv F_{1}(Q^{2}) + F_{2}(Q^{2})\,.
\label{flux_aux2}
\end{eqnarray}
Using the standard dipole parametrizations of the Sachs form factors
(see, for instance, chapter~2 in \cite{Close:2007zzd})
\begin{eqnarray}
&&G_{E}(Q^{2}) = G_{D}(Q^{2})\,,\qquad
G_{M}(Q^{2}) = \frac{\mu_{p}}{\mu_{N}} G_{D}(Q^{2})\,,
\label{dipole_form}\\  
&&G_{D}(Q^{2}) = \left( 1+\frac{Q^{2}}{m_{D}^{2}} \right)^{-2}\,, \; m_{D}^{2} = 0.71\,\mathrm{GeV}^{2}\,, 
\label{dipole}
\end{eqnarray}
where $G_{D}$ is the so-called dipole form factor, 
$\frac{\mu_{p}}{\mu_{N}} = 2.7928$,
$\mu_{p}$ and $\mu_{N}$ are the anomalous proton magnetic moment
and the nuclear magneton, respectively,
we obtain
%
\begin{eqnarray}
&&F_{1}(Q^{2}) = \left( 1+\frac{Q^{2}}{4m_{p}^{2}} \frac{\mu_{p}}{\mu_{N}} \right)
               \left( 1+\frac{Q^{2}}{4m_{p}^{2}} \right)^{-1}   G_{D}(Q^{2})\,,
\label{F1_dipole}\\
&&F_{2}(Q^{2}) = \left( \frac{\mu_{p}}{\mu_{N}} - 1 \right)
               \left( 1+\frac{Q^{2}}{4m_{p}^{2}} \right)^{-1}   G_{D}(Q^{2})\,.
\label{F2_dipole}
\end{eqnarray}
We shall use the parametrizations in the following analysis.

\subsection{Exact kinematics}

In the present studies we perform, for the first time, exact 
calculations for the considered exclusive $2 \to 4$ process (\ref{2to4_reaction}).
In general, the cross section 
can be written as
\begin{eqnarray}
d \sigma &=& 
\frac{(2 \pi)^{4}}{2s} {\overline{|{\cal M}_{pp \to pp H^+ H^-}|^2}}
\frac{d^3 p_1}{(2 \pi^{3})2 E_1} \frac{d^3 p_2}{(2 \pi^{3})2 E_2}
\frac{d^3 p_3}{(2 \pi^{3})2 E_3} \frac{d^3 p_4}{(2 \pi^{3})2 E_4} 
\nonumber \\
&&\times \delta^{4} \left(E_a + E_b -p_1 - p_2 - p_3 - p_4 \right)  \,,
\label{differential_cs}
\end{eqnarray}
where energy and momentum conservations have been made explicit.
The formula is written in the overall center-of-mass frame.
Above $\overline{|{\cal M}|^2}$ is the $2 \to 4$ amplitude squared 
averaged over initial and summed over final proton polarization states.
The kinematic variables for the reaction (\ref{2to4_reaction}) are
\begin{eqnarray}
&&s = (p_{a} + p_{b})^{2}, 
\quad s_{34} = M_{H^{+}H^{-}}^{2} = (p_{3} + p_{4})^{2},\nonumber \\
&& t_1 = q_{1}^{2},  \quad t_2 = q_{2}^{2}, 
\quad q_1 = p_{a} - p_{1},  \quad q_2 = p_{b} - p_{2}\,.
\label{2to4_kinematic}
\end{eqnarray}

Our calculations have been done using the VEGAS routine \cite{Lepage:1980dq}
and checked on an eight-dimensional grid
\footnote{The details on how to conveniently reduce the number of kinematic 
integration variables are discussed in \cite{Lebiedowicz:2009pj}.}.
The phase space integration variables are taken the same as 
in Ref.\cite{Lebiedowicz:2009pj}, except that proton transverse momenta 
$p_{1t}$ and $p_{2t}$ are replaced by
$\xi_1$ = log$_{10}(p_{1t}/p_{0t})$ and 
$\xi_2$ = log$_{10}(p_{2t}/p_{0t})$, respectively, where $p_{0t}$ = 1 GeV.
The main ingredients of the model are the amplitudes for the exclusive process. 

The Born amplitudes for the process (\ref{2to4_reaction}) are calculated as
\begin{eqnarray}
{\cal M}^{Born}_{\lambda_a \lambda_b \to \lambda_1 \lambda_2 H^+ H^-}(t_{1},t_{2}) =
V_{\lambda_a \to \lambda_1}^{\mu_1}(t_{1})
D_{\mu_1 \nu_1}(t_1) 
V_{\gamma \gamma \to H^+ H^-}^{\nu_1 \nu_2}
D_{\nu_2 \mu_2}(t_2)
V_{\lambda_b \to \lambda_2}^{\mu_2}(t_{2}) \,, \quad
\label{born}
\end{eqnarray}
where $D_{\mu \nu}(t) = -i g_{\mu \nu}/t$ is the photon propagator.
Using the Gordon 
decomposition the $\gamma pp$ vertex takes the form
%
\begin{eqnarray}
V_{\lambda \to \lambda'}^{(\gamma pp) \mu}(t) &=& e \, \bar{u}(p',\lambda') 
\left( 
\gamma^{\mu} F_{1}(t) + \frac{i}{2 m_{p}} \sigma^{\mu \nu} (p'-p)_{\nu} F_{2}(t)
\right) u(p,\lambda) \nonumber \\
&=& e \, \bar{u}(p',\lambda') 
\left( 
\left( F_{1}(t) + F_{2}(t) \right) \gamma^{\mu}  - \frac{\Bbb1}{2 m_{p}} (p'+p)^{\mu} F_{2}(t)
\right) u(p,\lambda) \,,
\label{vertex_spinors}
\end{eqnarray}
where $u(p,\lambda)$ is a Dirac spinor and $p, \lambda$ and $p', \lambda'$ are initial and final four-momenta and helicities of the protons, respectively.

In the high-energy approximation, at not too large $|t|$,
\footnote{
We show how good the approximation is in Figs. 9, 10, 12, 13.}
one gets the simple formula
%
\begin{eqnarray}
V_{\lambda \to \lambda'}^{(\gamma pp) \mu}(t) \simeq e
\left( \frac{\sqrt{-t}}{2 m_p} \right)^{| \lambda' - \lambda |} F_{i}(t) (p'+p)^{\mu} \,,
\label{vertex}
\end{eqnarray}
which is very convenient for the discussion of the proton spin-conserving 
and the proton spin-flipping 
components separately.
It is easy to see that in the approximation
[see Eq.~(\ref{vertex})] the cross section contains
only terms proportional to $F_{i}^{2}(t_{1})F_{j}^{2}(t_{2})$
and no mixed terms proportional to $F_{1}^{2}(t_{1})F_{1}(t_{2})F_{2}(t_{2})$, etc.
In exact calculations [with spinors of protons, see Eq.~(\ref{vertex_spinors})],
there is a small contribution of the mixed terms.
This will be discussed when presenting our results.

The tensorial vertex in Eq.~(\ref{born}) for the 
$\gamma \gamma \to H^+H^-$ subprocess
is a sum of three-level amplitudes corresponding to 
$t$, $u$ and contact diagrams of Fig.~\ref{fig:Born_diagrams}, respectively,
\begin{eqnarray}
V_{\gamma \gamma \to H^+ H^-}^{\nu_1 \nu_2} &=& V_t^{\nu_1 \nu_2} + V_u^{\nu_1 \nu_2} + V_c^{\nu_1 \nu_2} \nonumber \\
&=& i e^{2} \frac{1}{p_{t}^{2} - m_{H}^{2}} 
(q_{2} - p_{4} + p_{3})^{\nu_1} (q_{2} - 2p_{4})^{\nu_2}  \nonumber \\
&&+ i e^{2} \frac{1}{p_{u}^{2} - m_{H}^{2}}
(q_{1} - 2p_{4})^{\nu_1} (q_{1} - p_{4} + p_{3})^{\nu_2}  
  - 2 i e^{2} g^{\nu_1 \nu_2} \,,
\label{central_vertex}
\end{eqnarray}
where $p_{t}^{2} = (q_{2} - p_{4})^{2} = (q_{1} - p_{3})^{2}$
and $p_{u}^{2} = (q_{1} - p_{4})^{2} = (q_{2} - p_{3})^{2}$.
There are strong cancellations between the three contributions.

A complete calculation for exclusive $H^{+} H^{-}$ production
in $pp$ collisions, in addition to the $\gamma \gamma$ exchange,
must take into account more diagrams than those of Fig.~\ref{fig:Born_diagrams}.
We can have the $\gamma Z$, $Z \gamma$, and $ZZ$ exchanges.
The corresponding amplitudes can be obtained by substitution
of the photon propagator and the $\gamma pp$ vertex 
[see (\ref{born}) and (\ref{vertex_spinors})]
by the $Z$ boson propagator and the $Z pp$ vertex \cite{Alberico:2001sd},
\begin{eqnarray}
V_{\lambda \to \lambda'}^{(Zpp) \mu}(t) &=& \dfrac{e}{s_{W} c_{W}} \, \bar{u}(p',\lambda') 
\left( 
\gamma^{\mu} F_1^{NC}(t)
+ \frac{i}{2 m_{p}} \sigma^{\mu \nu} (p'-p)_{\nu} F_2^{NC}(t)
+ \gamma^{\mu} \gamma_5 G_A^{NC}(t)
\right) u(p,\lambda) \,,\nonumber \\
\label{vertex_Zpp}
\end{eqnarray}
where we use the shorthand notation $c_{W} \equiv \cos \theta_{W}$,
$s_{W} \equiv \sin \theta_{W}$, $\theta_{W}$ is the Weinberg mixing angle.
The $\gamma H^+ H^-$ and $\gamma \gamma H^+ H^-$ coupling constants 
in (\ref{central_vertex}) read:
\begin{eqnarray}
g_{Z H^+ H^-} &=& \dfrac{i e}{2 c_{W}s_{W}}(c_{W}^{2}-s_{W}^{2}) \,,\nonumber \\
g_{\gamma Z H^+ H^-} &=& \dfrac{i e^{2}}{c_{W}s_{W}}(c_{W}^{2}-s_{W}^{2}) \,,\nonumber \\
g_{Z Z H^+ H^-} &=& \dfrac{i e^{2}}{2 c_{W}^{2} s_{W}^{2}}(c_{W}^{2}-s_{W}^{2})^{2}\,.
\label{Zbosoncoupling}
\end{eqnarray}
The neutral current form factors appearing in (\ref{vertex_Zpp}) 
related to the vector part can be related to
electromagnetic form factors 
[see (\ref{F1_dipole}) and (\ref{F2_dipole}), $F_{1,2}(t) \equiv F_{1,2}^{p}(t)$],
\begin{equation}
F_{1,2}^{NC}(t) = \frac{1}{2} \left( F_{1,2}^{p}(t) -
  F_{1,2}^{n}(t) \right)
- 2 s^2_W F_{1,2}^{p}(t) - \frac{1}{2} F_{1,2}^{s}(t) \,,
\end{equation}
where $|F_{1}^{n}(t)| \ll |F_{1}^{p}(t)|$ for small $|t|$.
The form factor related to axial-vector neutral current
is related to the familiar charge current axial-vector form factor:
\begin{equation}
G_A^{NC}(t) = \frac{1}{2} G_A(t) - \frac{1}{2} G_A^{s}(t) \,.
\end{equation}
Since the strangeness form factors $F_{1,2}^{s}$ and $G_A^{s}$ are poorly known
and small in the following we shall neglect them.

\subsection{Absorption corrections}

The absorptive corrections to the Born amplitude (\ref{born})
are added to give the full physical amplitude for the $pp \to pp H^{+} H^{-}$ reaction: 
\begin{eqnarray}
{\cal {M}}_{pp \to pp H^{+} H^{-}} =
{\cal {M}}_{pp \to pp H^{+} H^{-}}^{\mathrm{Born}} + 
{\cal {M}}_{pp \to pp H^{+} H^{-}}^{\mathrm{absorption}}\,.
\label{amp_full}
\end{eqnarray}
Here (and above) we have for simplicity omitted 
the dependence of the amplitude on kinematic variables.


The amplitude including $pp$-rescattering corrections 
between the initial- and final-state protons
in the four-body reaction discussed here can be written as
%
\begin{eqnarray}
{\cal M}_{\lambda_a \lambda_b \to \lambda_1 \lambda_2 H^+ H^-}^{\mathrm{absorption}}(s,\bp_{1t},\bp_{2t})=&&
\frac{i}{8 \pi^{2} s} \int d^{2} \bk_{t} 
{\cal M}_{\lambda_{a}\lambda_{b} \to \lambda'_{a}\lambda'_{b}}(s,-{\bk}_{t}^{2})\nonumber \\
&&\times
{\cal M}_{\lambda'_{a}\lambda'_{b}\to \lambda_{1}\lambda_{2} H^{+} H^{-}}^{\mathrm{Born}}(s,\tilde{\bp}_{1t},\tilde{\bp}_{2t})\,, \nonumber \\
\label{abs_correction}
\end{eqnarray}
where $\tilde{\bp}_{1t} = {\bp}_{1t} - {\bk}_{t}$ and
$\tilde{\bp}_{2t} = {\bp}_{2t} + {\bk}_{t}$.
Here, in the overall center-of-mass system, ${\bp}_{1t}$ and ${\bp}_{2t}$
are the transverse components of the momenta of the final-state protons
and ${\bk}_{t}$ is the transverse momentum carried by additional pomeron exchange.
${\cal M}_{pp \to pp}(s,-{\bk}_{t}^{2})$ 
is the elastic $pp$-scattering amplitude
for large $s$ and with the momentum transfer $t=-{\bk}_{t}^{2}$.
Here we assume $s$-channel helicity conservation
and the exponential functional form of form factors 
in the pomeron-proton-proton vertices.

We shall show results in the Born approximation
as well as include the absorption corrections on the amplitude level.
This allows us to study the absorption effects differentially in any
kinematical variable chosen, which has, so far, never been done for two-photon induced (sub)processes.
\section{Results}
\label{sec:section_3}
\subsection{Electromagnetic process}
\label{sec:electromagnetic_process}

In this section we shall present results of our calculations
for the $p p \to p p H^+ H^-$ reaction (\ref{2to4_reaction})
calculating from the diagrams of Fig.~\ref{fig:Born_diagrams}.
Let us start our presentation by presenting the total cross section
for $\sqrt{s}$ = 14 TeV (LHC) and $\sqrt{s}$ = 100 TeV (FCC)
and for various charged Higgs mass values.
In Table~\ref{table:cross_sections} we show 
cross sections in fb without and with (results in the parentheses) 
the $pp$-rescattering corrections.
The smaller the values of $m_{H^{\pm}}$, the larger are those of cross section
\footnote{We wish to note on the margin that the cross section for 
pair production for doubly charged (Higgs) bosons,
e.g. $H^{++} H^{--}$, would be 16 times larger 
\cite{Han:2007bk,Kanemura:2014goa,Kanemura:2014ipa} in the leading-order
approximation considered here.
The doubly charged Higgs bosons are expected in models that contain a Higgs boson triplet field.}.
The values of the gap survival factor $\langle S^{2}\rangle$
for different masses of $H^{\pm}$ bosons $m_{H^{\pm}} = 150, 300, 500$~GeV
are, respectively, $0.77, 0.67, 0.57$ for $\sqrt{s} = 14$~TeV (LHC)
and $0.89, 0.86, 0.82$ for $\sqrt{s} = 100$~TeV (FCC).
In contrast to diffractive processes, the larger the collision energy, 
the smaller the effect of absorption.
We have checked numerically that the cross section contributions
with the $\gamma Z$, $Z \gamma$, and $ZZ$ exchanges
are very small compared to the $\gamma \gamma$ contribution
and will be not presented explicitly in this paper.
\begin{table}
\begin{tabular}{|l|c|c|c|}
\hline
$m_{H^{\pm}}$ (GeV)   &   150    & 300     & 500 \\   
\hline
$\sigma_{LHC}$ (fb)   &   0.1474 (0.1132) &  0.0119 (0.0080) &  0.0014 (0.0008) \\
$\sigma_{FCC}$ (fb)   &   1.0350 (0.9236) &  0.1470 (0.1258) &  0.0303 (0.0249) \\
\hline
\end{tabular}
\caption{ \small
Cross sections in fb for the $p p \to p p H^+ H^-$ reaction through photon-photon exchanges
without and with (results in the parentheses) the absorption corrections
for two center-of-mass energies $\sqrt{s} = 14$~TeV (LHC) and $\sqrt{s} = 100$~TeV (FCC)
and various charged Higgs bosons mass values. 
The calculations was performed for exact $2 \to 4$ kinematics 
and with the amplitudes in the high-energy approximation, see Eq.~(\ref{vertex}).
}
\label{table:cross_sections}
\end{table}

\begin{figure}
\includegraphics[width=0.48\textwidth]{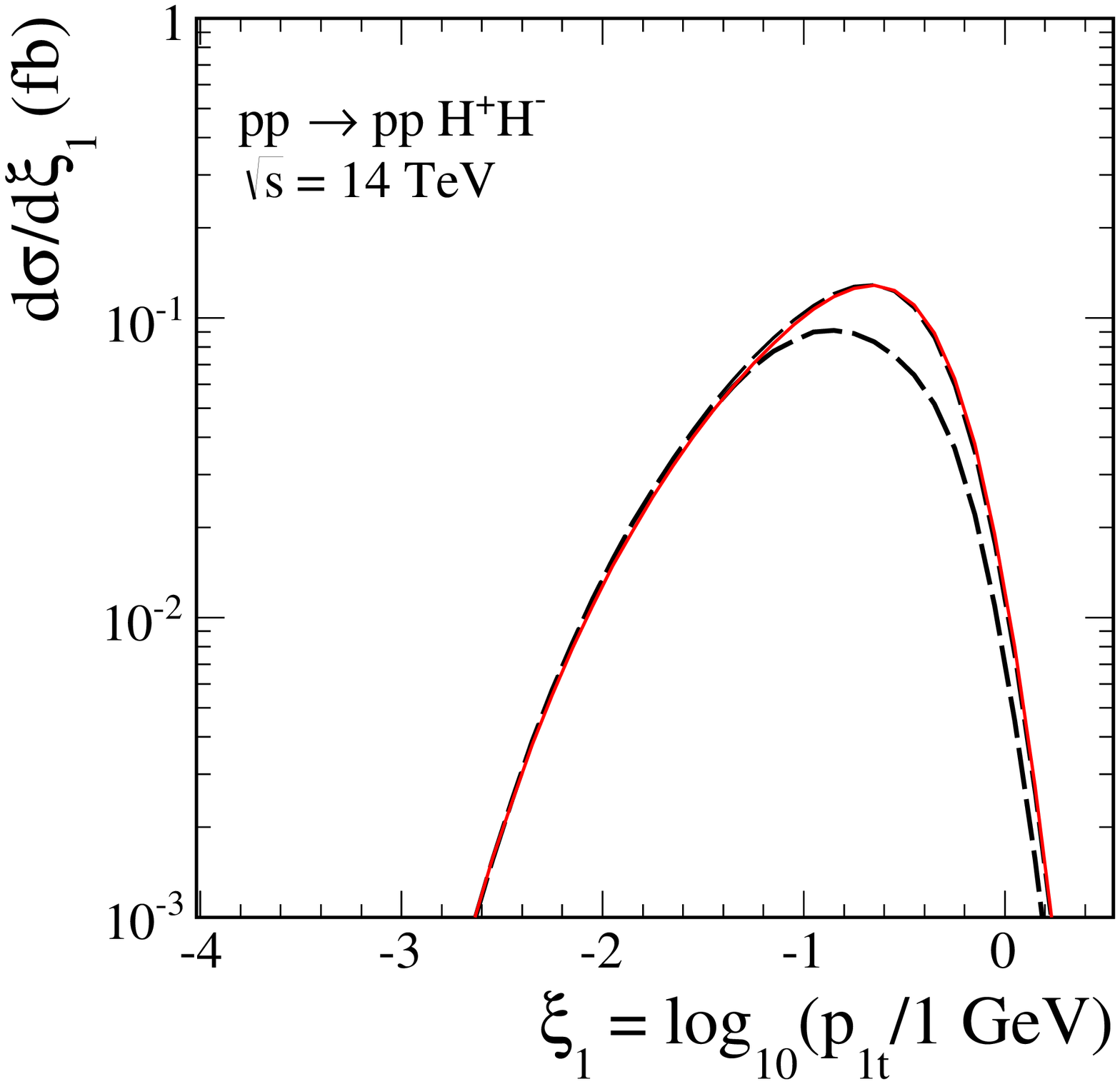}
\includegraphics[width=0.48\textwidth]{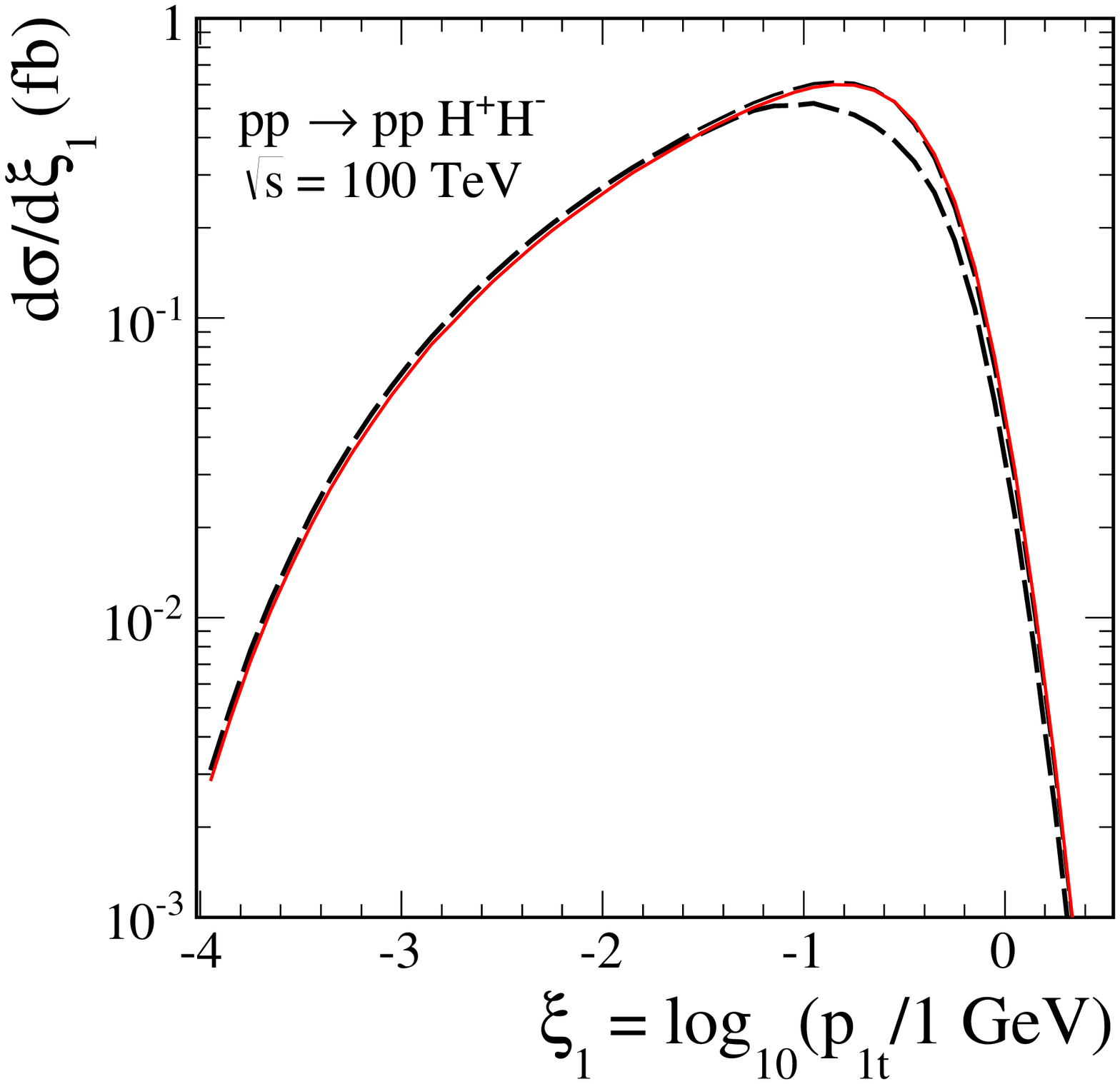}
\caption{Distribution in the auxiliary variables $\xi_1$ or $\xi_2$
at $\sqrt{s} = 14$~TeV (left panel) and $100$~TeV (right panel).
The online red solid lines represent the calculation of exact amplitude 
(including spinors of protons).
The black upper and lower long-dashed lines correspond to calculations 
in the high-energy approximation (\ref{vertex})
without and with the absorption corrections, respectively.
}
\label{fig:dsig_dxi}
\end{figure}

In Fig.~\ref{fig:dsig_dxi} we show a distribution in an auxiliary
integration variable(s) $\xi_{1/2}$ = $\log_{10}(p_{1/2t}/1\,\mathrm{GeV})$.
If protons are measured, the distributions in Fig.~\ref{fig:dsig_dxi} can be measured too.
Here and in the following, we discuss the differential distributions
for one selected mass of $H^{\pm}$.
For example, we shall assume $m_{H^{\pm}} = 150$~GeV, which is 
rather a lower limit for the charged Higgs bosons. 
The general features of the differential distribution for heavier masses are, however, similar.
We compare results without (the upper long-dashed lines) 
and with (the lower long-dashed lines) 
absorption corrections due to the $pp$ interactions. 

\begin{figure}
\includegraphics[width=0.48\textwidth]{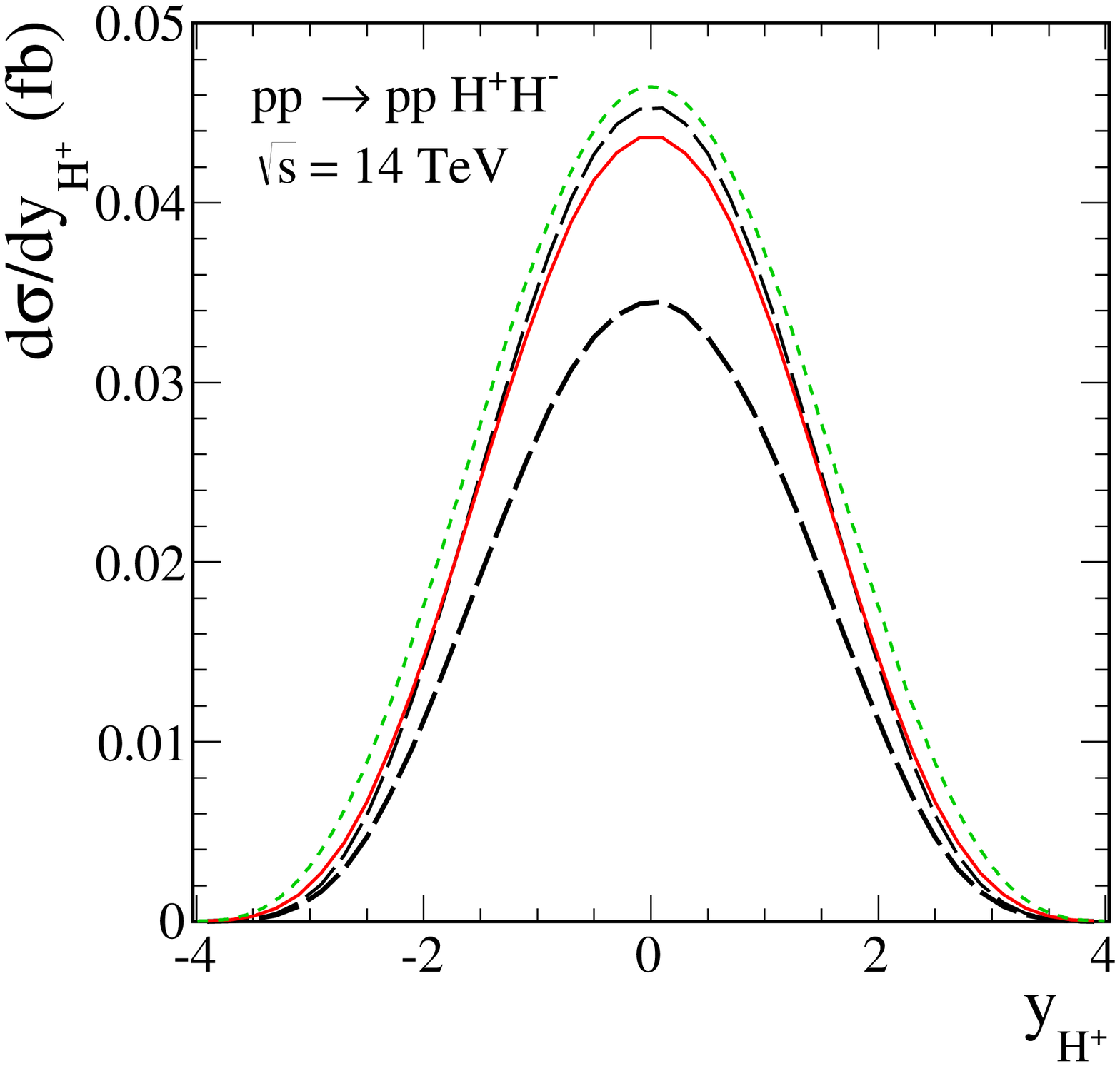}
\includegraphics[width=0.48\textwidth]{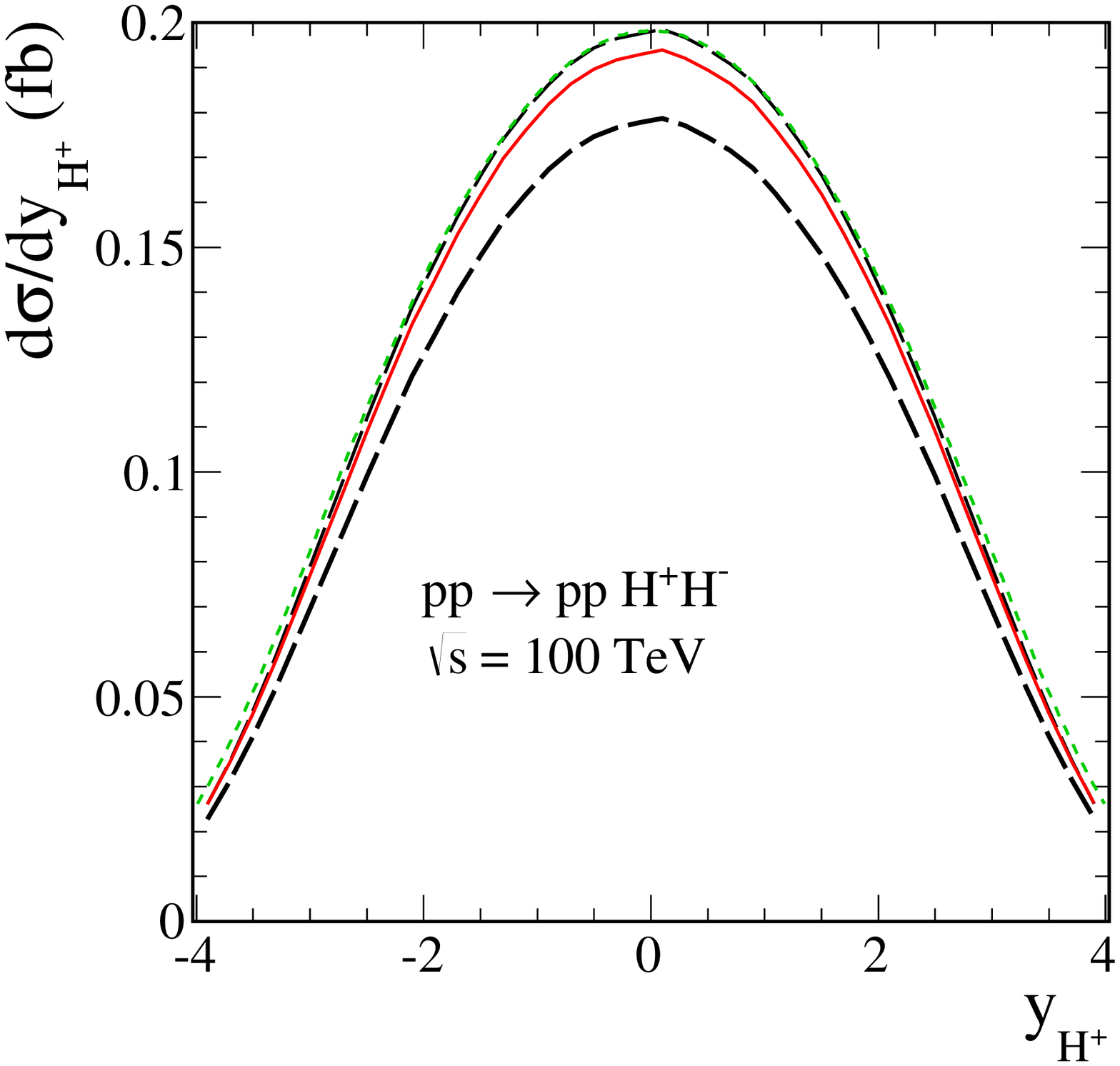}
\caption{Rapidity distribution of charged (Higgs) bosons
at $\sqrt{s} = 14$~TeV (left panel) and $100$~TeV (right panel).
The meaning of the lines is the same as in Fig.~\ref{fig:dsig_dxi}.
The short-dashed (online green) lines represent results of EPA.
} 
\label{fig:dsig_dyH}
\end{figure}

The rapidity distribution for the charged Higgs bosons is shown in Fig.~\ref{fig:dsig_dyH}. 
The larger center-of-mass energy the broader the rapidity distributions. 

\begin{figure}
\includegraphics[width=0.48\textwidth]{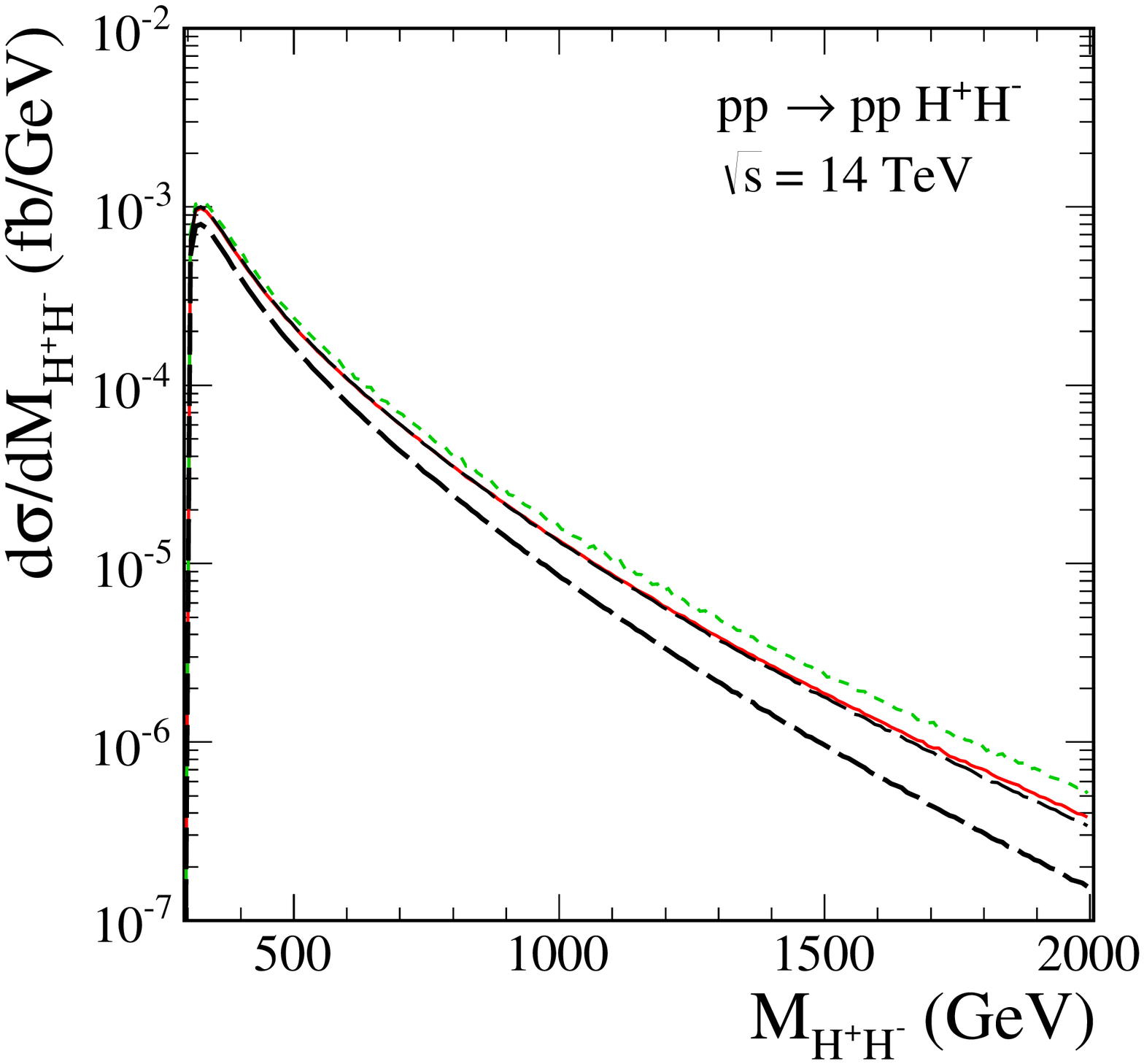}
\includegraphics[width=0.48\textwidth]{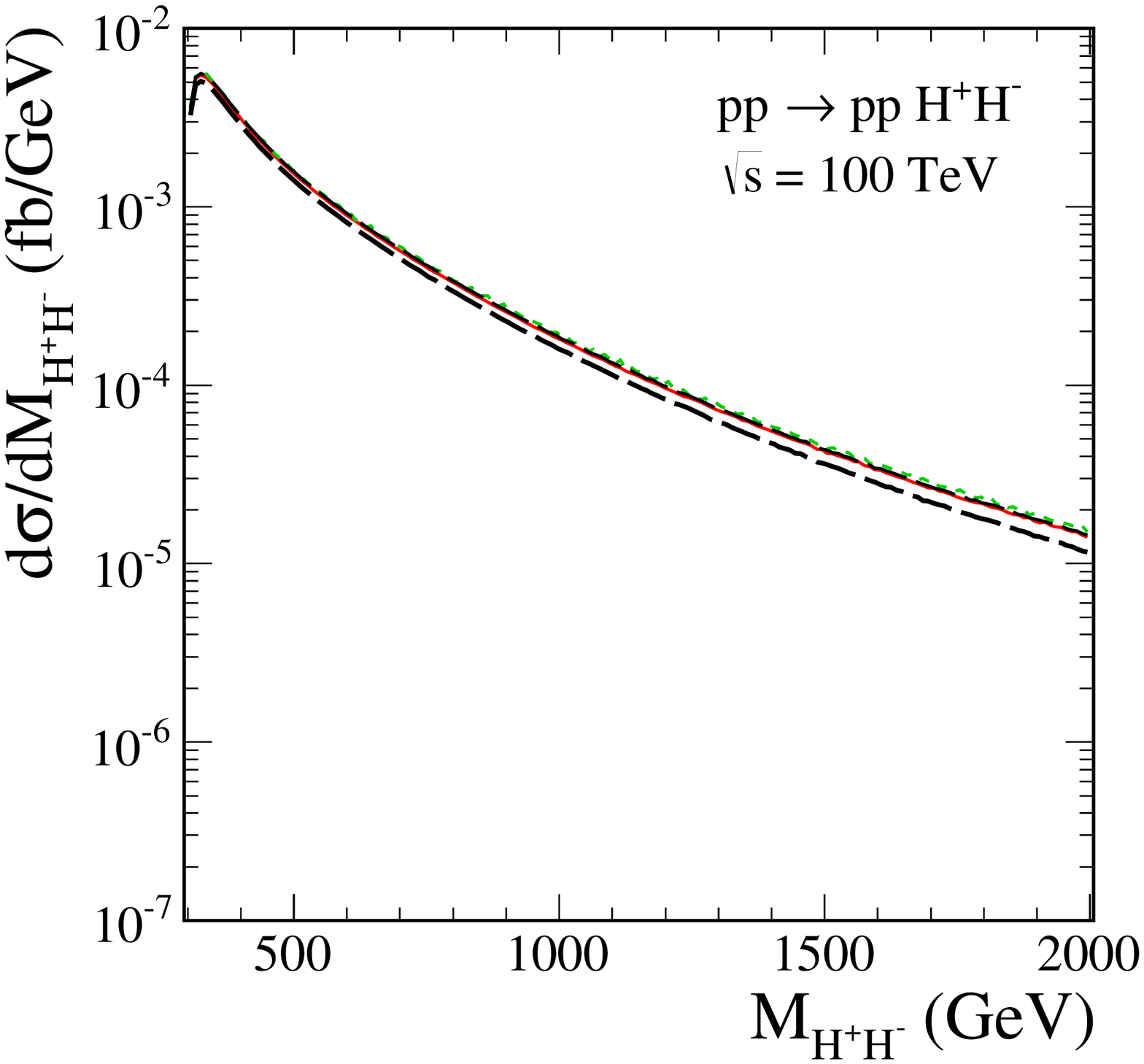}
\caption{
DiHiggs boson invariant mass distributions 
at $\sqrt{s} = 14$~TeV (left panel) and $100$~TeV (right panel).
The meaning of the lines is the same as in Fig.~\ref{fig:dsig_dxi}.
The short-dashed (online green) lines represent results of EPA.
}
\label{fig:dsig_dMHH}
\end{figure}

In Fig.~\ref{fig:dsig_dMHH} we show invariant mass distribution
of the $H^+ H^-$ subsystem in a broad range of the invariant masses.
We compare results for the exact kinematics and for the EPA calculations.
Please note that for the EPA,
the invariant mass of the diHiggs system is given by 
$M_{H^{+}H^{-}} \approx s x_{1} x_{2}$.

\begin{figure}
\includegraphics[width=0.48\textwidth]{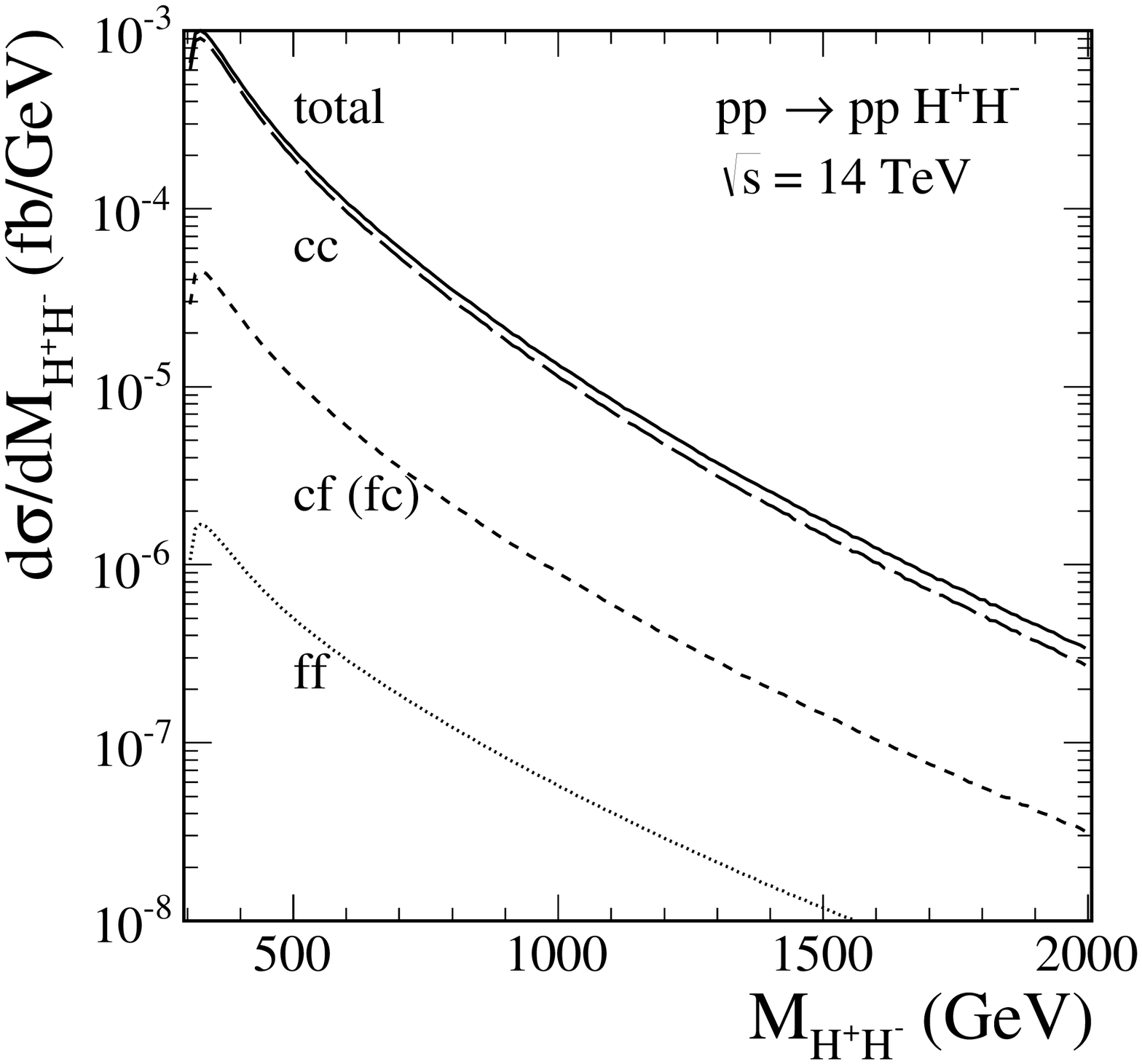}
\includegraphics[width=0.48\textwidth]{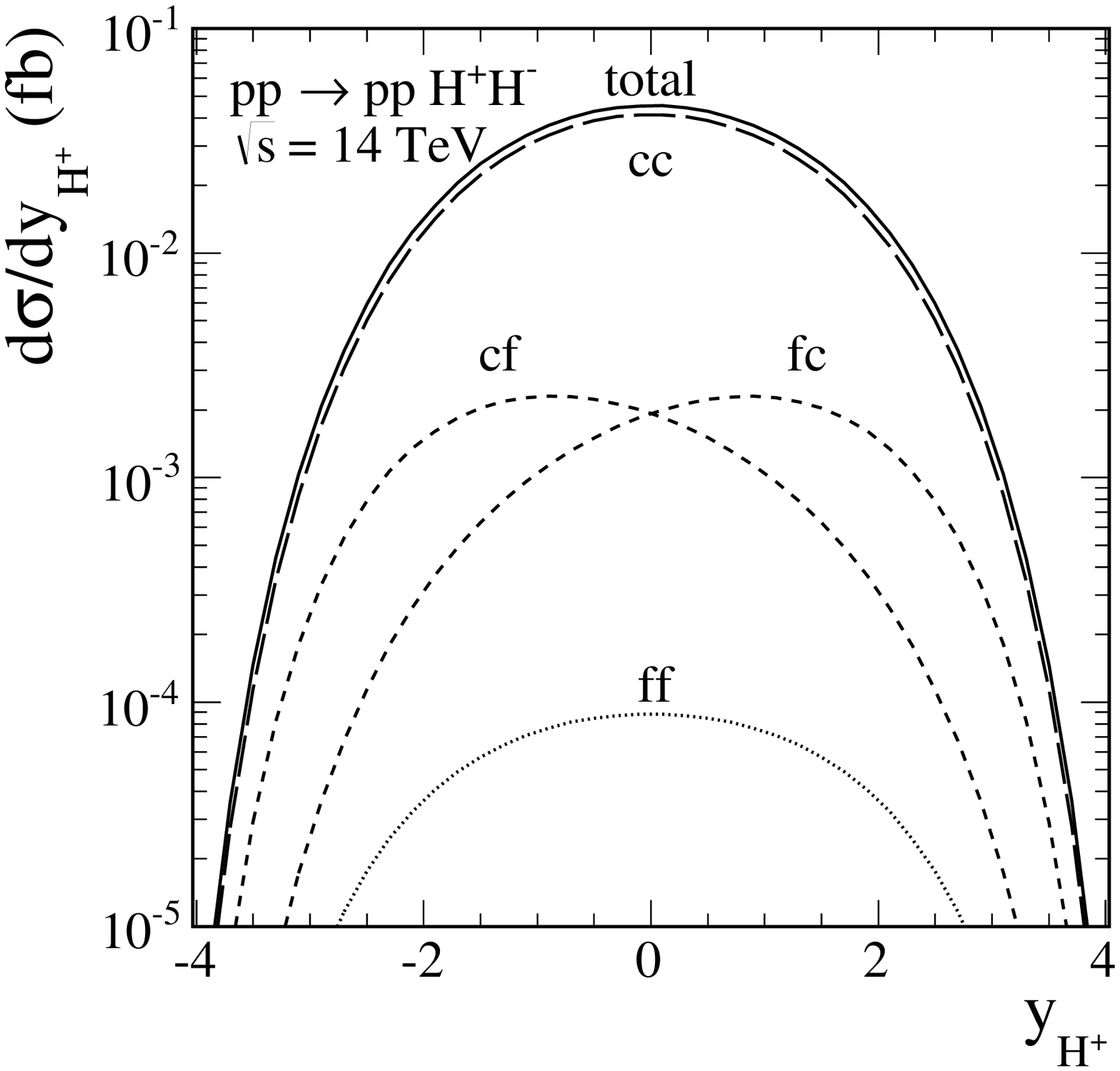}
\caption{
DiHiggs boson invariant mass (left panel) and Higgs boson rapidity 
(right panel) distributions for $\sqrt{s} = 14$~TeV 
in the Born approximation and for the amplitudes given by Eq.~(\ref{vertex}).
The double spin-conserving contribution (cc) is show by the long-dashed line,
the double spin-flipping contribution (ff) by the dotted line,
and the mixed contributions (cf) and (fc) by the dashed lines.
The solid line represents the sum of all the contributions
and corresponds to the upper long-dashed lines in 
Figs.~\ref{fig:dsig_dyH} and \ref{fig:dsig_dMHH}~(left panels).
}
\label{fig:dsig_deco}
\end{figure}

In Fig.~\ref{fig:dsig_deco} 
we show decomposition into helicity components of the cross section 
in the two-Higgs invariant mass
and in the rapidity of one of the charged Higgs bosons.
Here we use the formula (\ref{vertex}) for the $\gamma pp$ vertex
which is very convenient for the discussion of the proton spin-conserving
(the Dirac form factor (\ref{F1_dipole}) only)
and the proton spin-flipping 
(the Pauli form factor (\ref{F2_dipole}) only) components separately.

\begin{figure}[!ht]  
\center
\includegraphics[width = 0.48\textwidth]{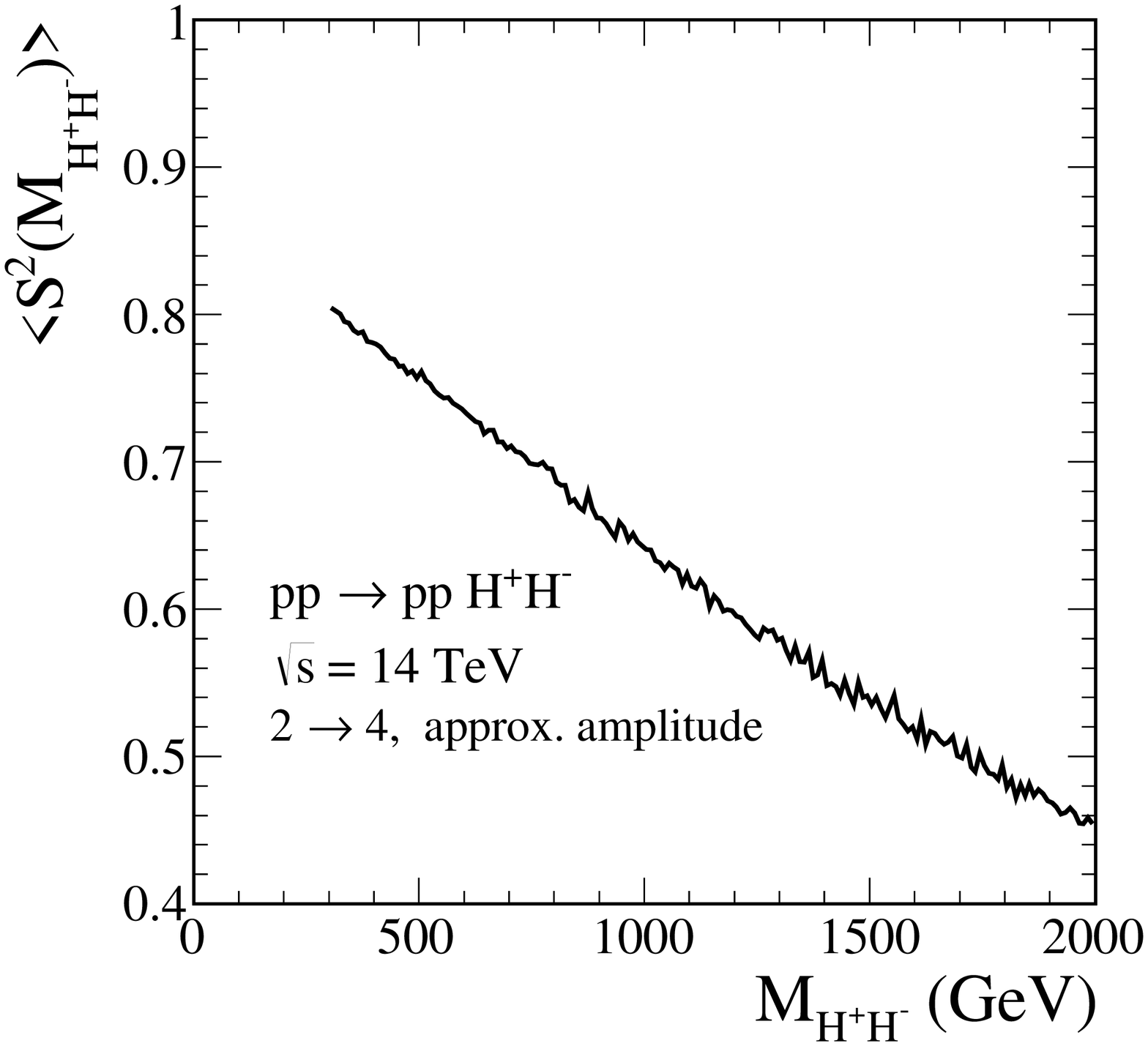}
\includegraphics[width = 0.48\textwidth]{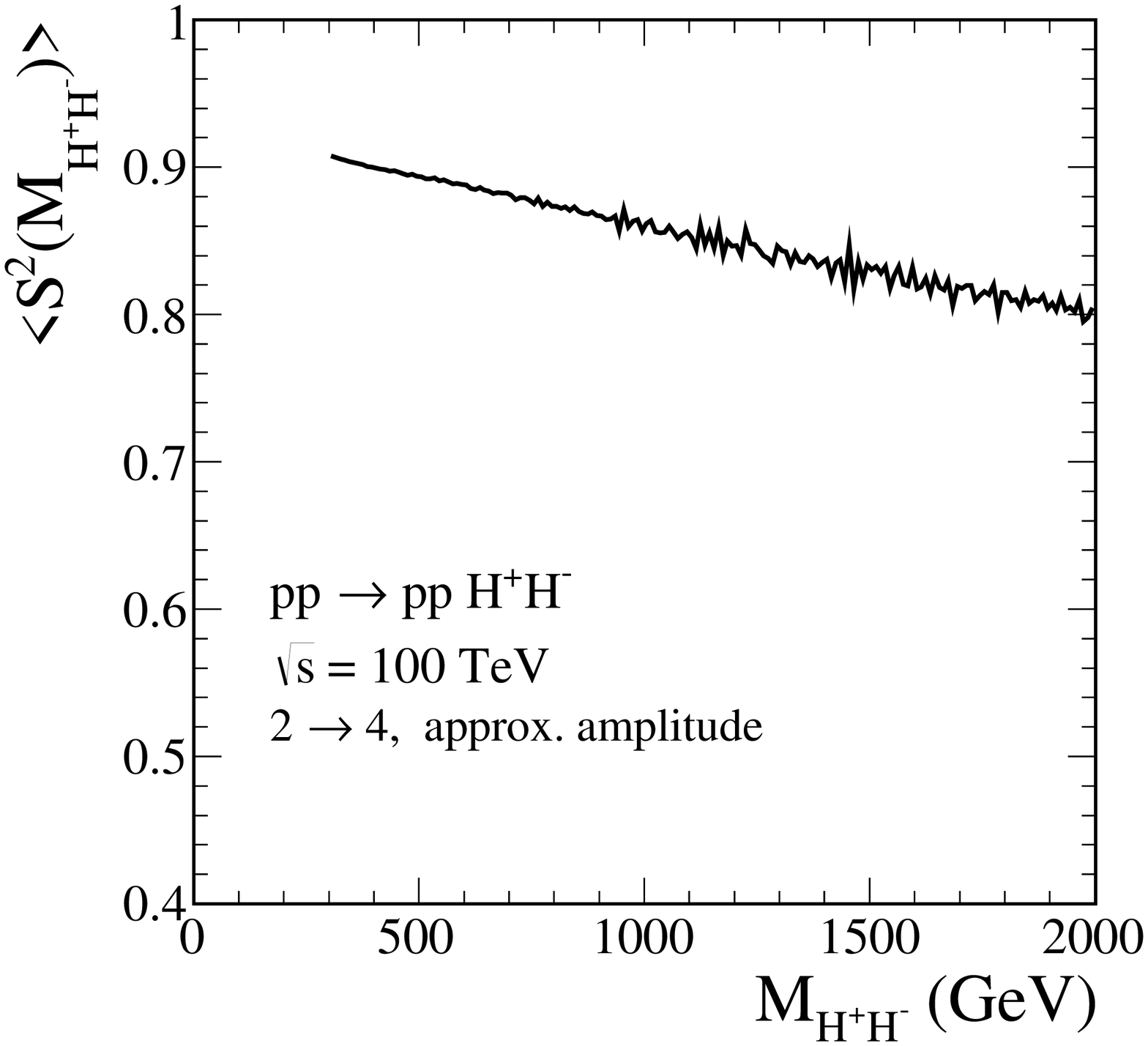}
  \caption{\label{fig:ratio_abs}
  \small
The dependence of the gap survival factor due to $pp$ interactions 
on $M_{H^{+}H^{-}}$ for exact $2 \to 4$ kinematics
at $\sqrt{s} = 14$~TeV (left panel) and 100~TeV (right panel).
This is quantified by the ratio of full (including absorption) 
and Born differential cross sections
$\langle S^{2}(M_{H^{+}H^{-}})\rangle$ (\ref{ratio_abs}).
}
\end{figure}

In Fig.~\ref{fig:ratio_abs} we show the dependence of absorption on $M_{H^{+}H^{-}}$.
This is quantified by the ratio of full (with the absorption corrections) and Born 
differential cross sections
\begin{eqnarray}
\langle S^{2}(M_{H^{+}H^{-}})\rangle = 
\frac{d\sigma^{\mathrm{Born\,+\,absorption}}/dM_{H^{+}H^{-}}}{d\sigma^{Born}/dM_{H^{+}H^{-}}} \,.
\label{ratio_abs}
\end{eqnarray}
The absorption effects due to the $pp$ interaction lead to large damping
of the cross section at the LHC and relatively small reduction of 
the cross section at the FCC. This result must be contrasted with typical
diffractive exclusive processes where the role of absorption effects
gradually increases with the collision energy.

\begin{figure}
\includegraphics[width=0.48\textwidth]{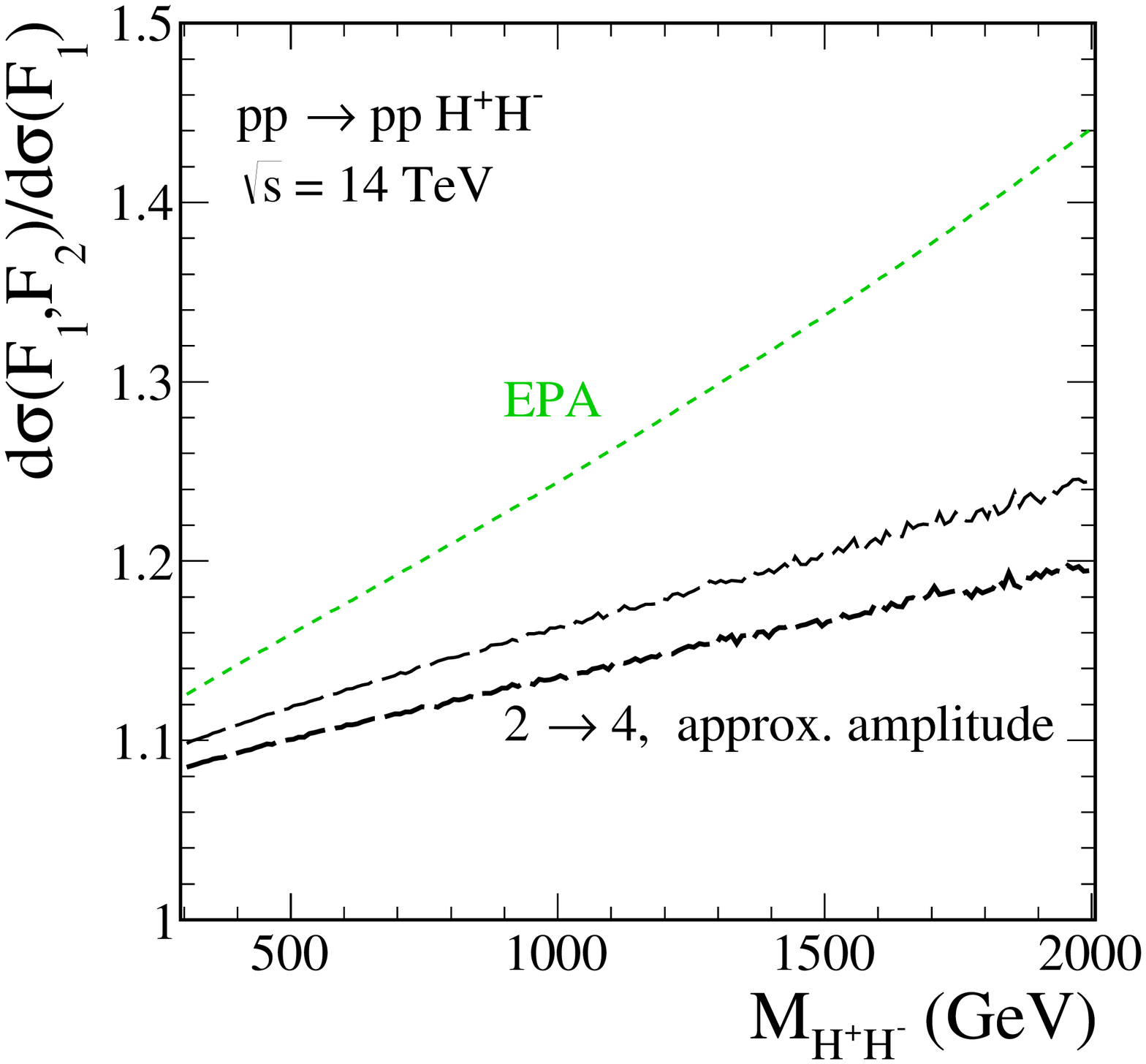}
\includegraphics[width=0.48\textwidth]{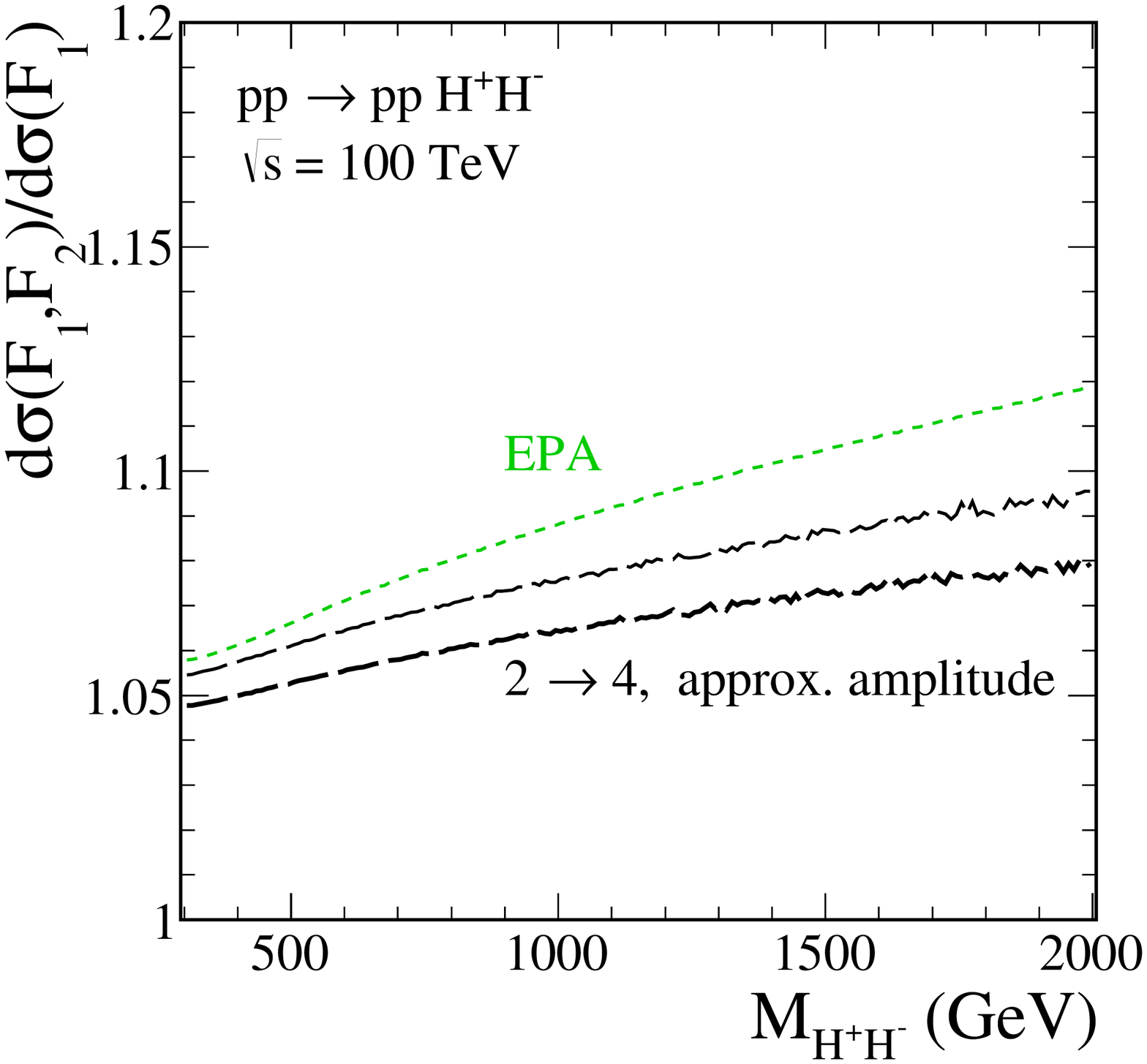}
\caption{Ratio of the cross section when all ($F_1$, $F_2$) terms are included
to that with $F_1$ terms only for exact $2 \to 4$ kinematics (black long-dashed lines) 
and EPA (online green short-dashed lines) calculations
at $\sqrt{s} = 14$~TeV (left panel) and $\sqrt{s} = 100$~TeV (right panel).
The lowest long-dashed lines represent results with the absorption corrections.
}
\label{fig:ratio_ff}
\end{figure}

In Fig.~\ref{fig:ratio_ff} we show the ratio of the cross section
for all ($F_1$, $F_2$) terms included in the amplitude to that for $F_1$ terms only 
both for the exact $2 \to 4$ kinematics and for the EPA calculations.
Here for consistency we have neglected the interference effect 
between the electromagnetic form factors in the EPA approach.
At large invariant masses of $M_{H^{+}H^{-}}$ the ratio for exact calculation 
is much smaller than that for EPA. 
This suggests that EPA overestimates the spin-flipping contributions.

\begin{figure}
\includegraphics[width=0.48\textwidth]{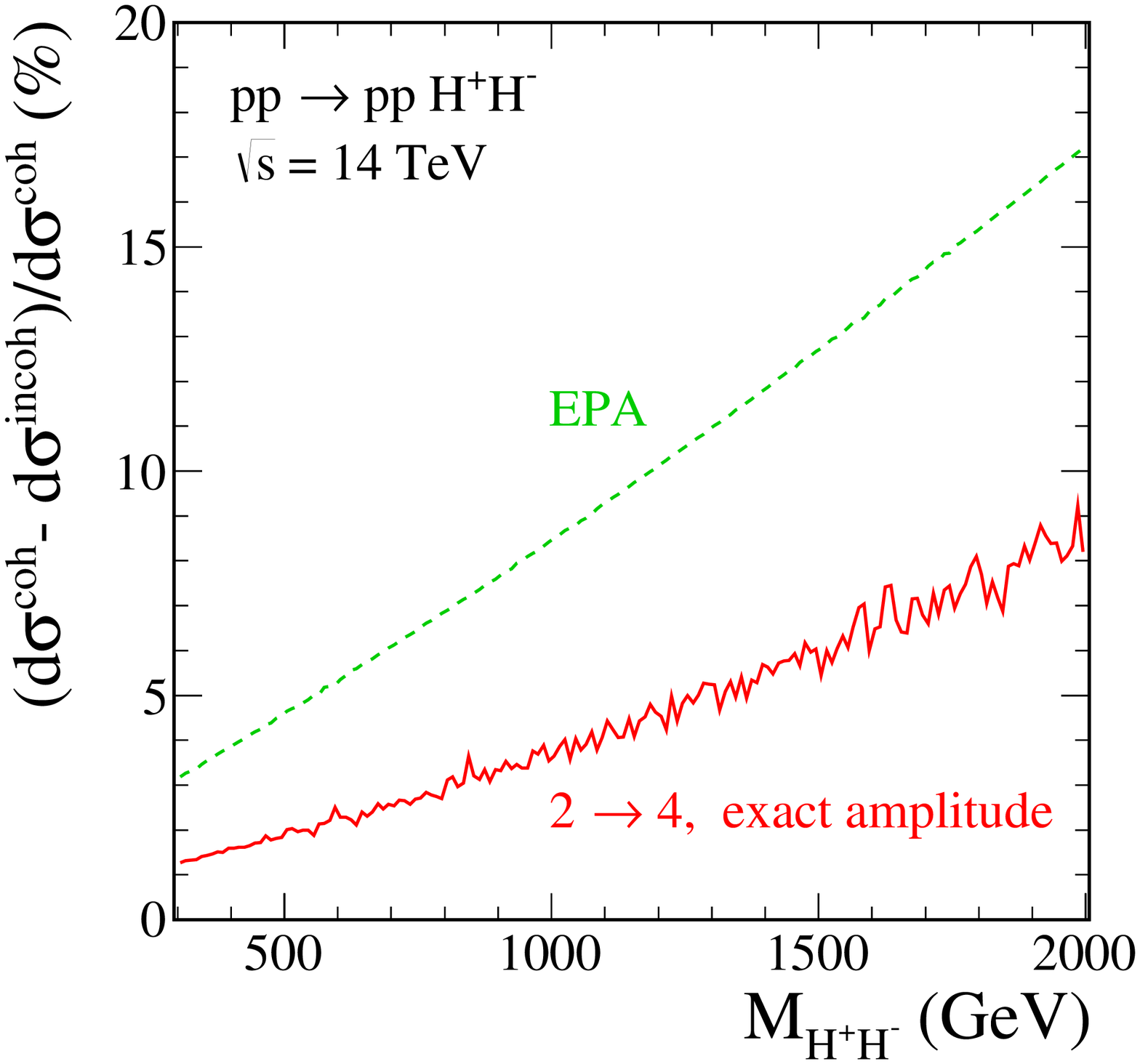}
\includegraphics[width=0.48\textwidth]{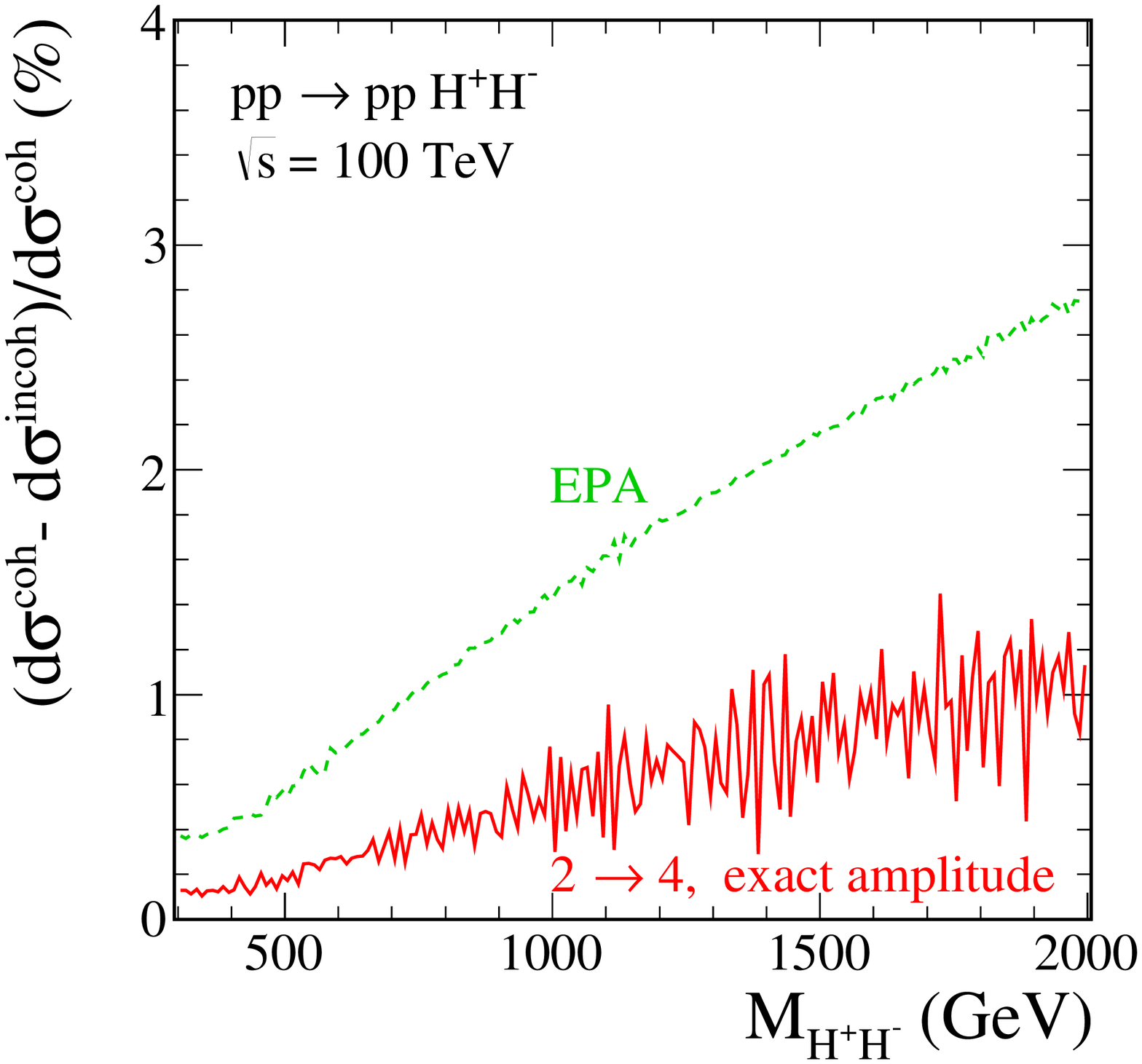}
\caption{The relative corrections in percent as a function of $M_{H^{+}H^{-}}$
for exact Born amplitude including spinors of protons (online red solid lines) 
and for EPA (online green short-dashed lines) calculations
at $\sqrt{s} = 14$~TeV (left panel) and $100$~TeV (right panel). 
}
\label{fig:ratio_interf}
\end{figure}

Let us discuss now a subtle effect of the interference of terms 
proportional to $F_{1}$ and $F_{2}$.
To quantify the effect, let us define the following quantities:
\begin{eqnarray}
&&d\sigma^{incoh} = d\sigma(F_{1};F_{1})+d\sigma(F_{1};F_{2})
                + d\sigma(F_{2};F_{1})+d\sigma(F_{2};F_{2})\,,\\
&&d\sigma^{coh} = d\sigma(F_{1},F_{2};F_{1},F_{2}) \,,
\label{xcross}
\end{eqnarray}
where $d\sigma(F_{i};F_{j})$ means the cross section when, at one proton line,
only the $F_{i}$ term is taken into account and, at the second proton line,
only the $F_{j}$ term is taken into account.
$d\sigma^{coh}$ represents the cross section where all terms are coherently included.
In Fig.~\ref{fig:ratio_interf} we show the relative corrections 
($(d\sigma^{coh}-d\sigma^{incoh})/d\sigma^{coh}$)
coming from the interference effect between different terms in the amplitude.
We see from Fig.~\ref{fig:ratio_interf} that for $M_{H^{+}H^{-}} = 2$~TeV
the total cross section from the calculation using exact amplitude 
(including spinors of protons) is modified by $\approx 10 \%$ at $\sqrt{s} = 14$~TeV, 
while at $\sqrt{s} = 100$~TeV only by $\approx 1 \%$.
The smallness of the effect causes the effect of the fluctuations
in our Monte Carlo approach.
The relative corrections for the EPA approach are somewhat larger.

\begin{figure}
\includegraphics[width=0.48\textwidth]{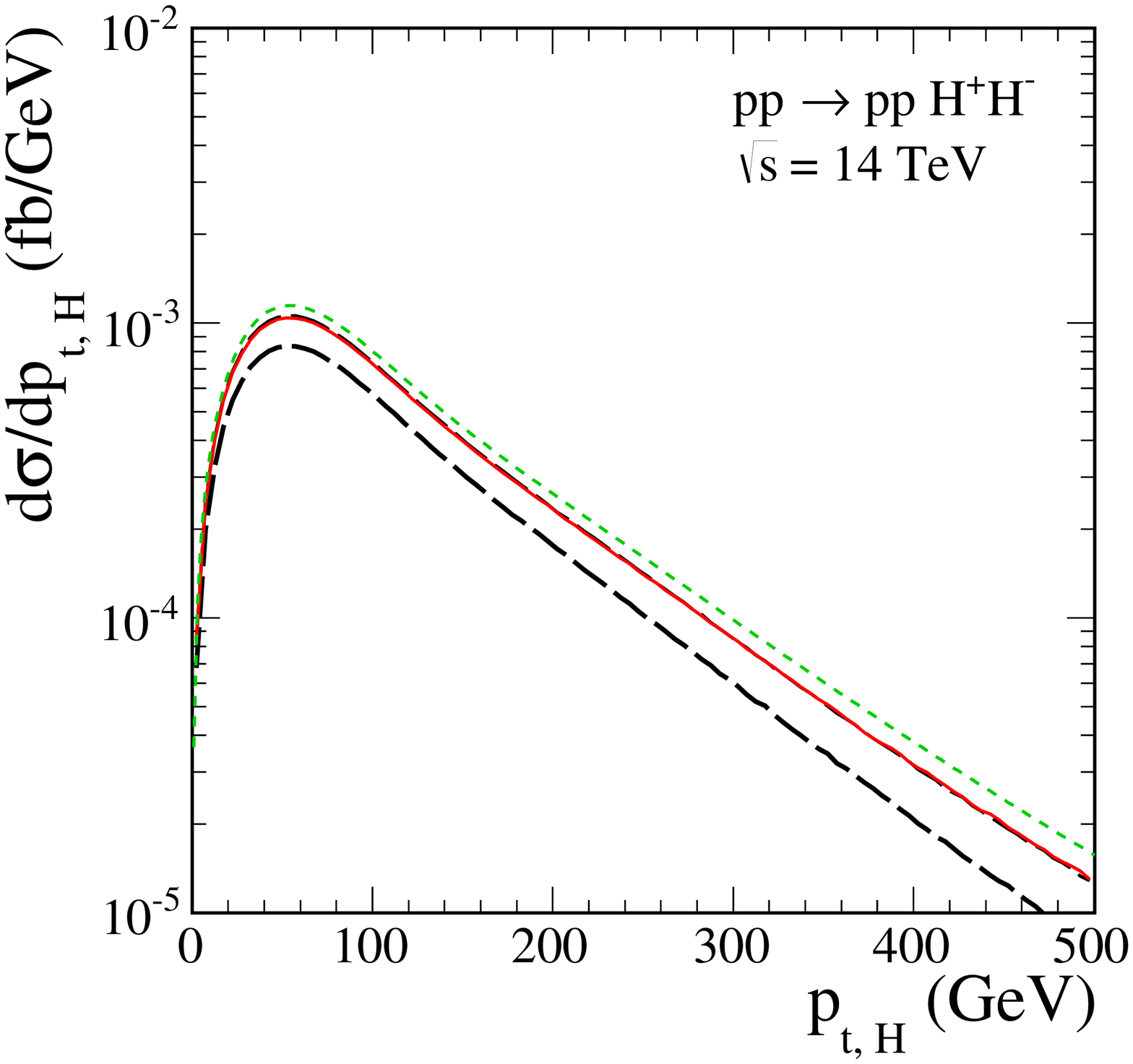}
\includegraphics[width=0.48\textwidth]{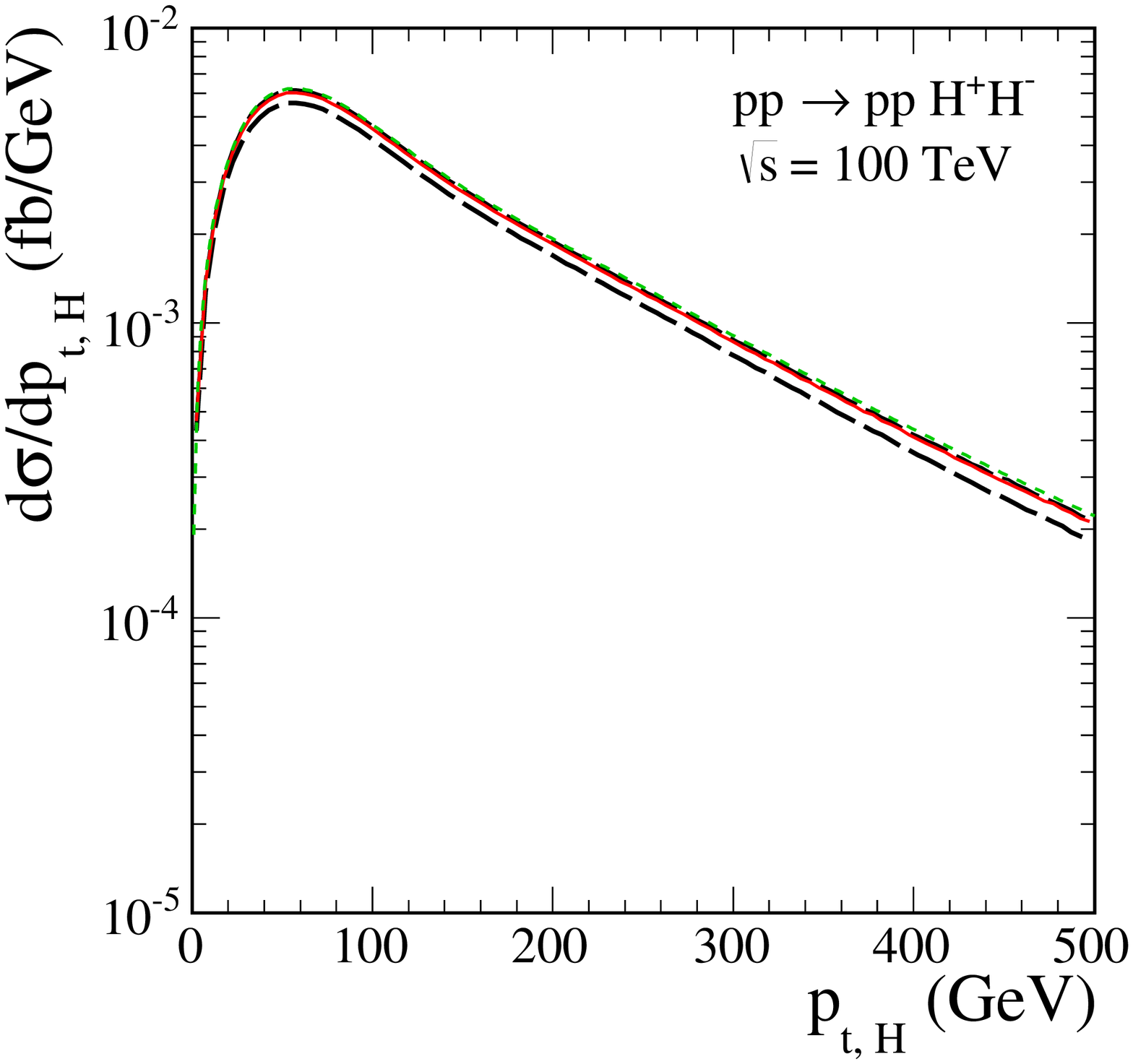}
\caption{The distributions in the Higgs boson transverse momentum
at $\sqrt{s} = 14$~TeV (left panel) and $100$~TeV (right panel).
The meaning of the lines is the same as in Fig.~\ref{fig:dsig_dxi}.
The online green short-dashed lines represent results of EPA.
}
\label{fig:dsig_dptH}
\end{figure}

In Fig.~\ref{fig:dsig_dptH} we present distributions
in charged Higgs boson transverse momentum $p_{t,H}$,
i.e., $p_{t,H^{+}}$ or $p_{t,H^{-}}$.
While at low (Higgs) boson transverse momenta the EPA result
is very similar to our exact result for all spin components, 
some deviations can be observed at larger transverse momenta. 
This is consistent with the similar comparison for the
distributions in invariant mass (for the process under consideration 
large transverse momenta are related to large invariant masses).

\begin{figure}
\includegraphics[width=0.48\textwidth]{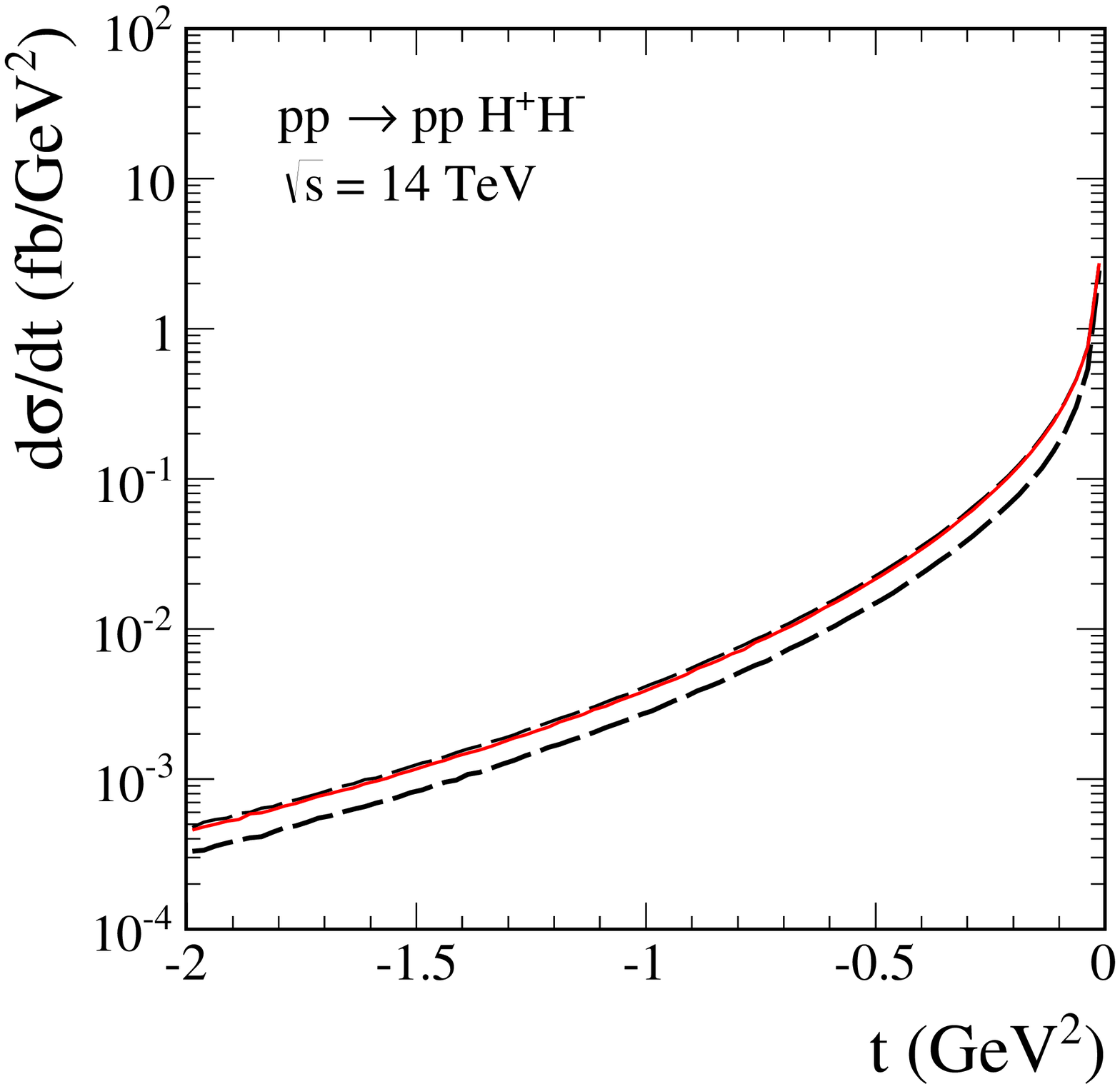}
\includegraphics[width=0.48\textwidth]{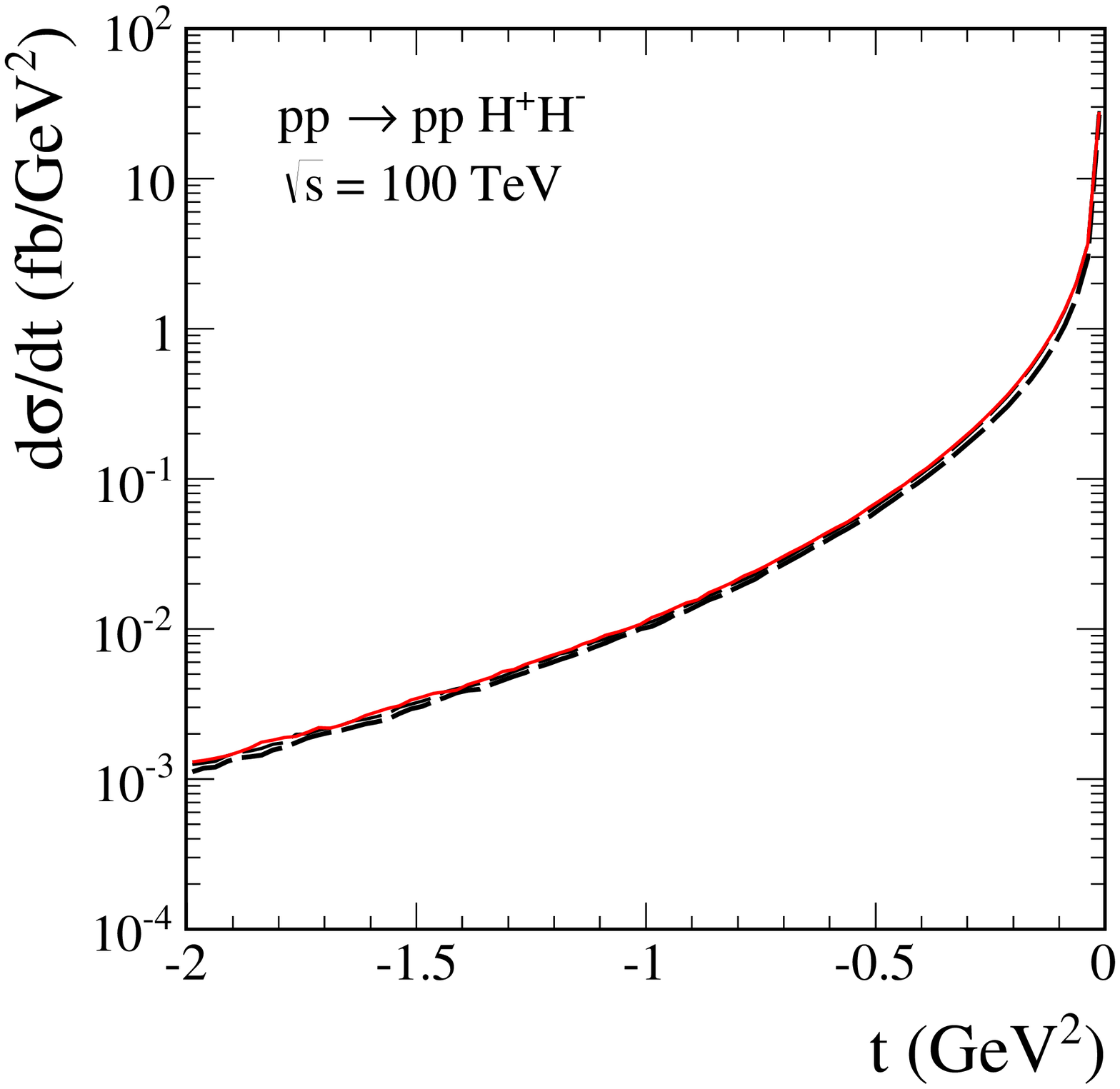}
\caption{Distribution in momentum transfer(s) squared ($t_1$ or $t_2$)
at $\sqrt{s} = 14$~TeV (left panel) and $100$~TeV (right panel).
The meaning of the lines is the same as in Fig.~\ref{fig:dsig_dxi}.
}
\label{fig:dsig_dt}
\end{figure}

If forward/backward protons are measured, then distributions
in four-momentum transfers squared ($t$ = $t_{1}$ or $t_{2}$) can be obtained
and relevant distributions shown in Fig.~\ref{fig:dsig_dt} can be constructed.
The absorption effects due to the $pp$ interactions are stronger for large values of $|t|$.

\begin{figure}
\includegraphics[width=0.48\textwidth]{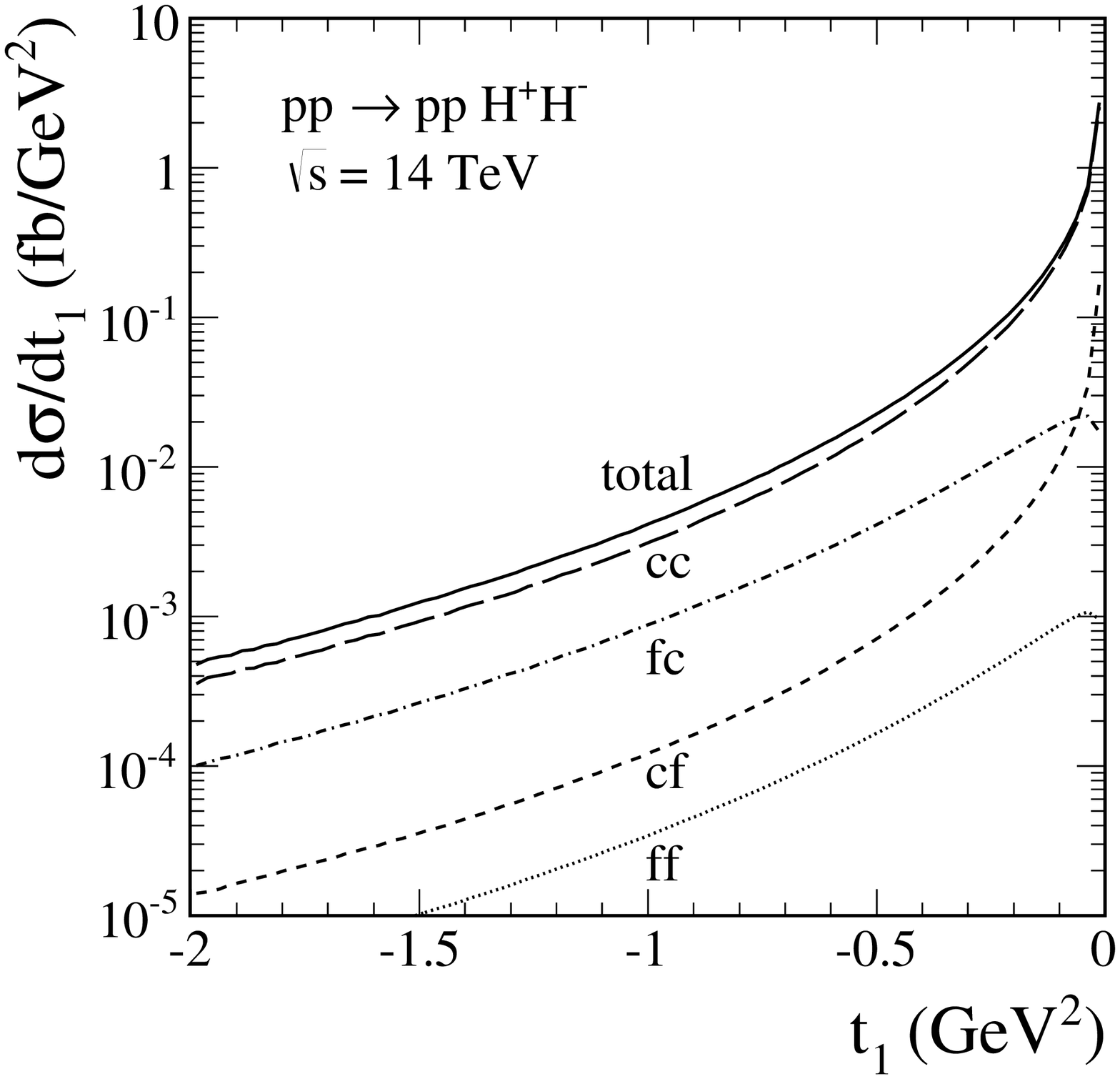}
\includegraphics[width=0.48\textwidth]{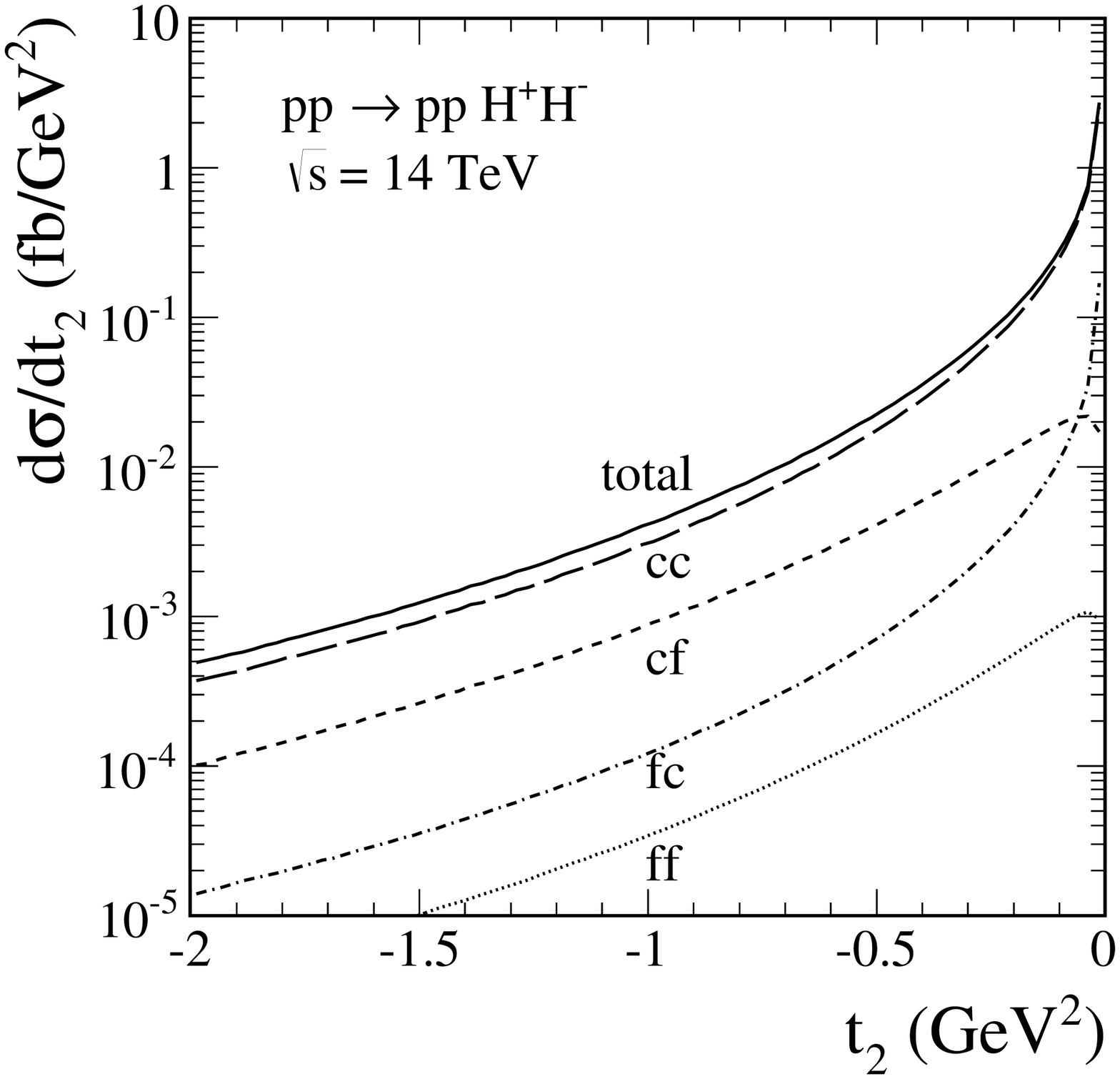}
\caption{Distributions in momentum transfer squared $t_{1}$ (left panel) 
and $t_{2}$ (right panel) at $\sqrt{s} = 14$~TeV in the Born
approximation and for the amplitudes given by Eq.~(\ref{vertex}).
The double spin-conserving contribution (cc) is show by the long-dashed line,
the double spin-flipping contribution (ff) by the dotted line,
and the mixed contributions (cf) and (fc) by the dashed and dot-dashed line, respectively.
The solid line represents the sum of all the contributions
and corresponds to the upper long-dashed line in Fig.~\ref{fig:dsig_dt}.
}
\label{fig:dsig_dt_deco}
\end{figure}

In Fig.~\ref{fig:dsig_dt_deco} we show a decomposition of the cross
section into helicity components as a function of momentum transfer(s) 
squared. The proton spin-conserving contribution related to the Dirac form
factor(s) clearly dominates at very small $|t_1|$ or $|t_2|$.
At larger $|t|$ the proton spin-flipping contribution related to 
the Pauli form factor(s) becomes important as well. 
The double spin-flipping contribution (ff) vanish at $|t_{1}| = |t_{2}| = 0$,
while the mixed contributions (fc) and (cf) vanish at $|t_{1}| = 0$ and 
$|t_{2}| = 0$, respectively.

\begin{figure}
\includegraphics[width=0.48\textwidth]{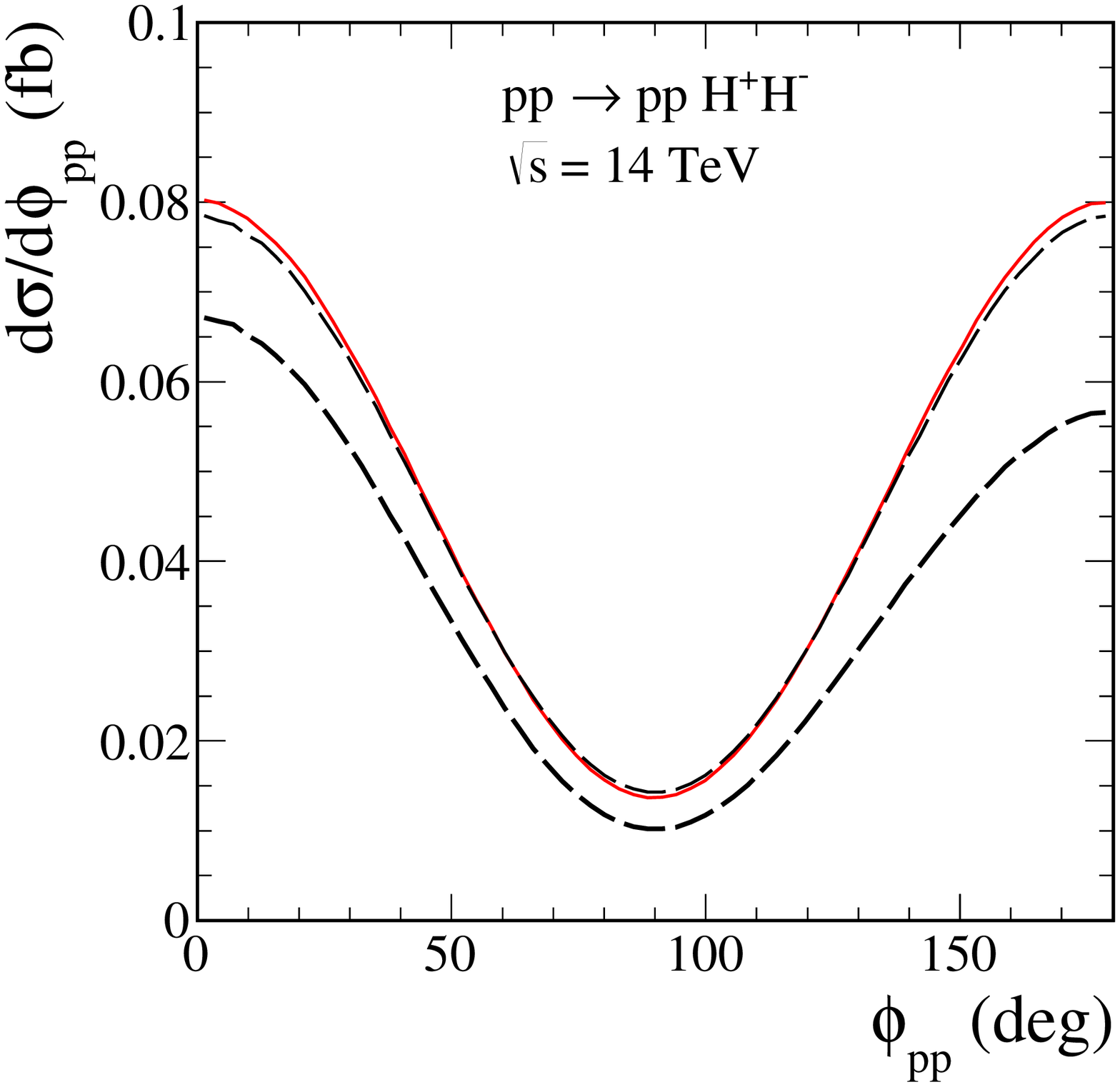}
\includegraphics[width=0.48\textwidth]{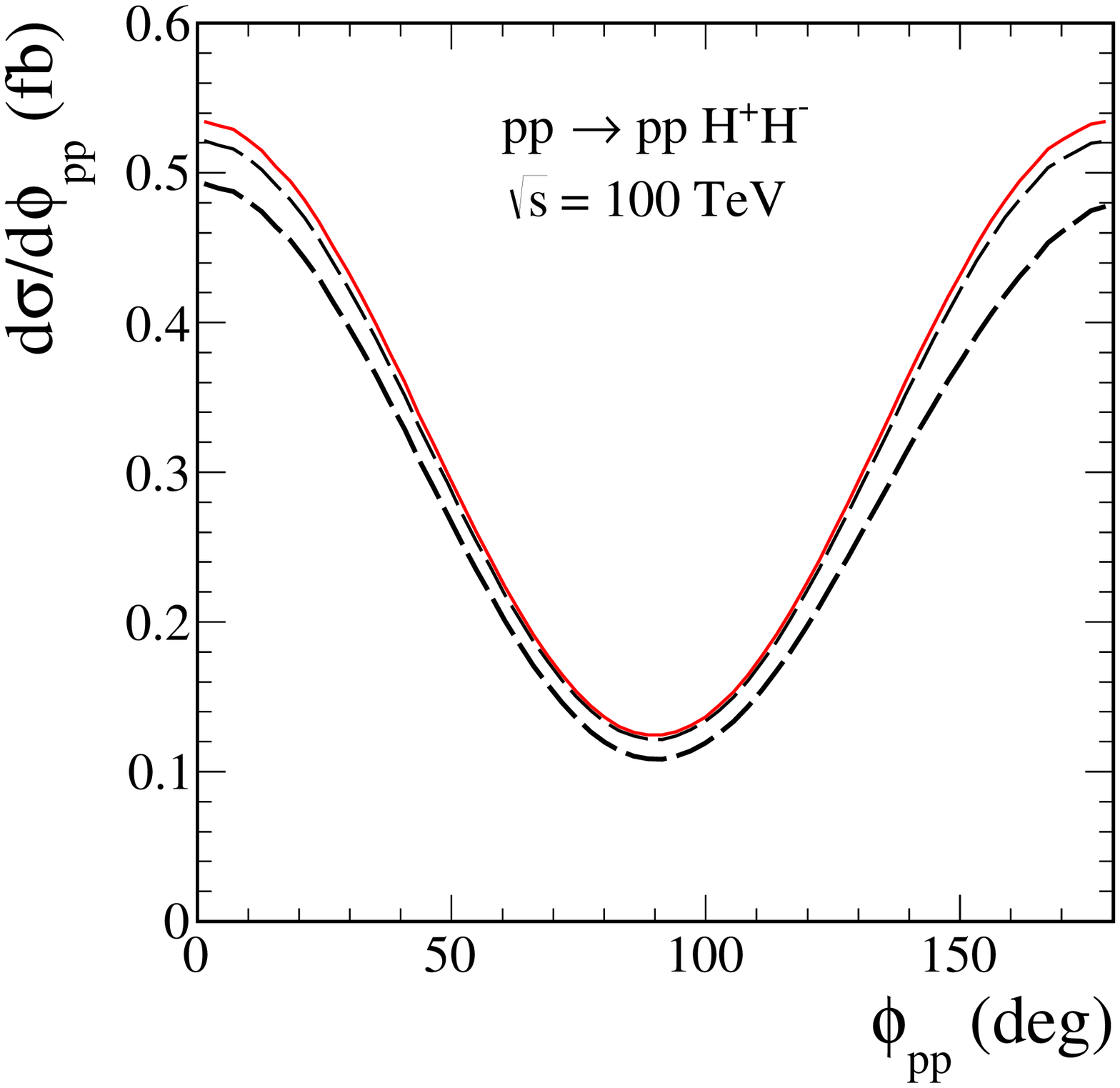}
\caption{Distribution in relative azimuthal angle between outgoing protons 
at $\sqrt{s} = 14$~TeV (left panel) and $100$~TeV (right panel).
The meaning of the lines is the same as in Fig.~\ref{fig:dsig_dxi}.}
\label{fig:dsig_dphipp}
\end{figure}

Let us consider now azimuthal correlations between outgoing particles.
In Fig.~\ref{fig:dsig_dphipp} we show correlations between outgoing protons.
We emphasize the dip at $\phi_{pp} = \pi/2$ which is a consequence of
the couplings involved in calculating the $\gamma \gamma \to H^+ H^-$ 
matrix element(s).

\begin{figure}
\includegraphics[width=0.48\textwidth]{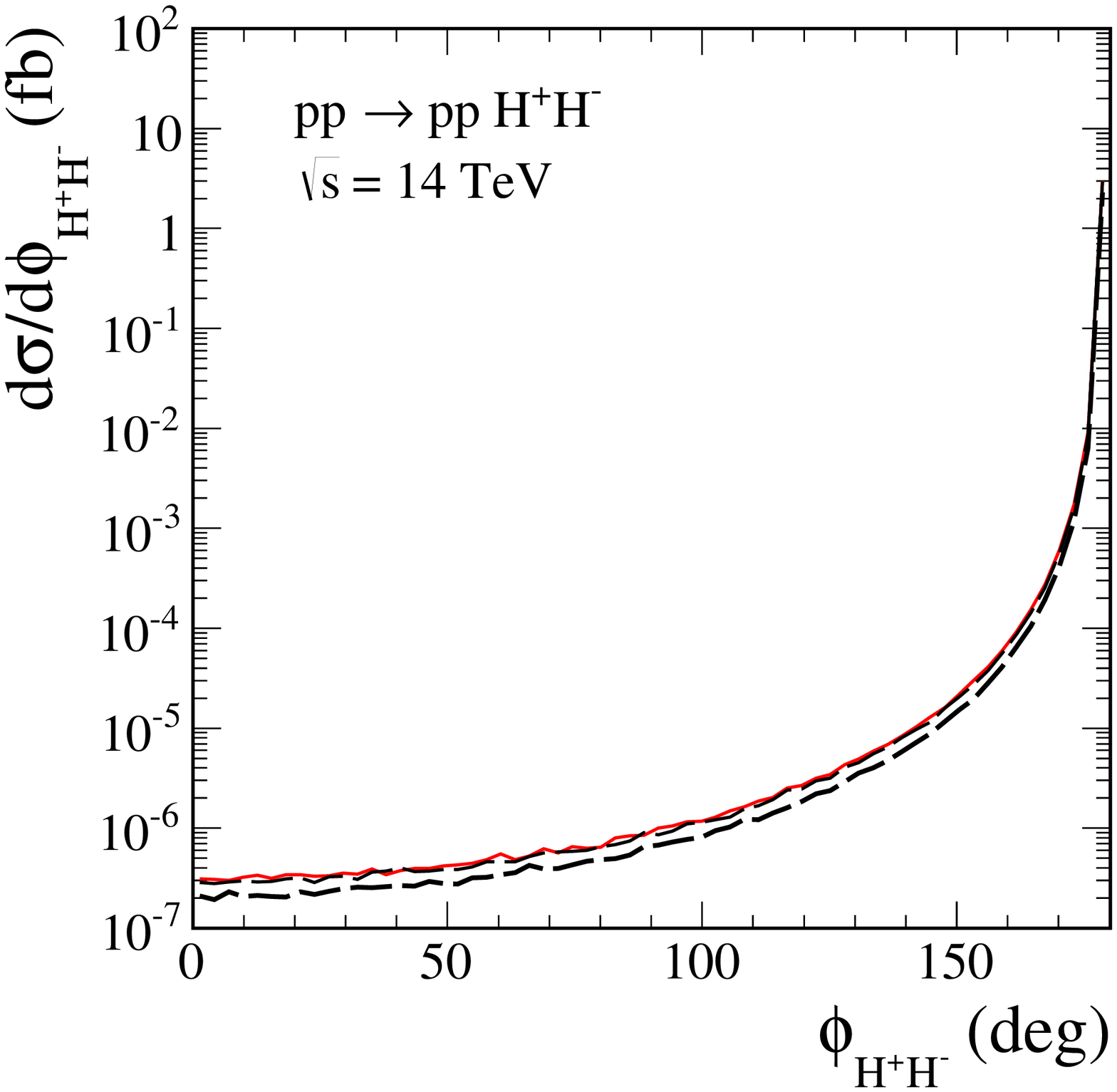}
\includegraphics[width=0.48\textwidth]{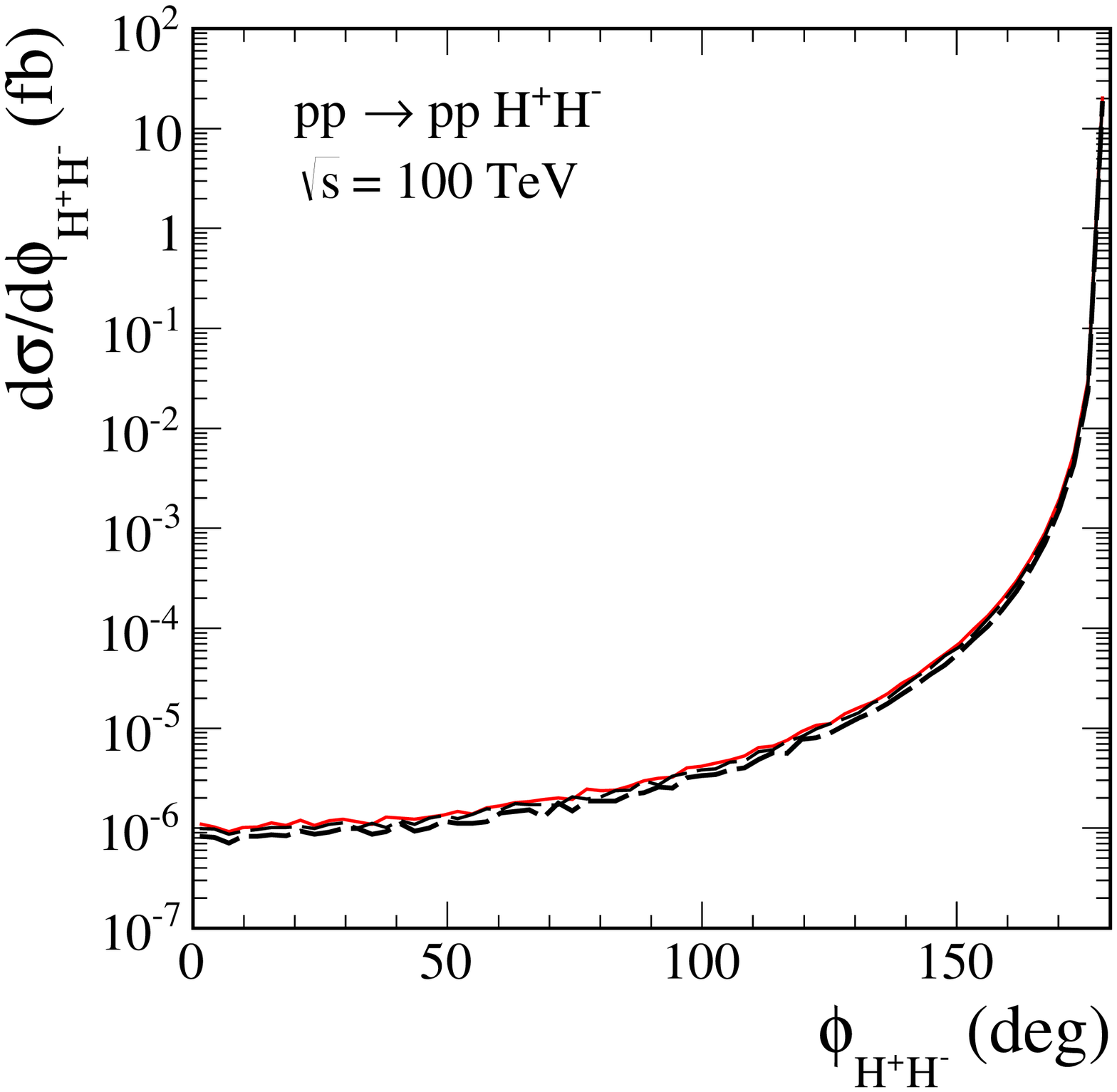}
\caption{Distribution in relative azimuthal angle between outgoing charged (Higgs) bosons
at $\sqrt{s} = 14$~TeV (left panel) and $100$~TeV (right panel).
The meaning of the lines is the same as in Fig.~\ref{fig:dsig_dxi}.
}
\label{fig:dsig_dphiHH}
\end{figure}

The correlation between outgoing Higgs bosons is shown in 
Fig.~\ref{fig:dsig_dphiHH}. The bosons are produced preferentially
back-to-back which can be understood given small transverse momenta
of virtual photons compared to transverse momenta of the Higgs bosons.

\subsection{Diffractive process}
\label{sec:diffractive_process}

\begin{figure}[!ht]
\includegraphics[width=0.4\textwidth]{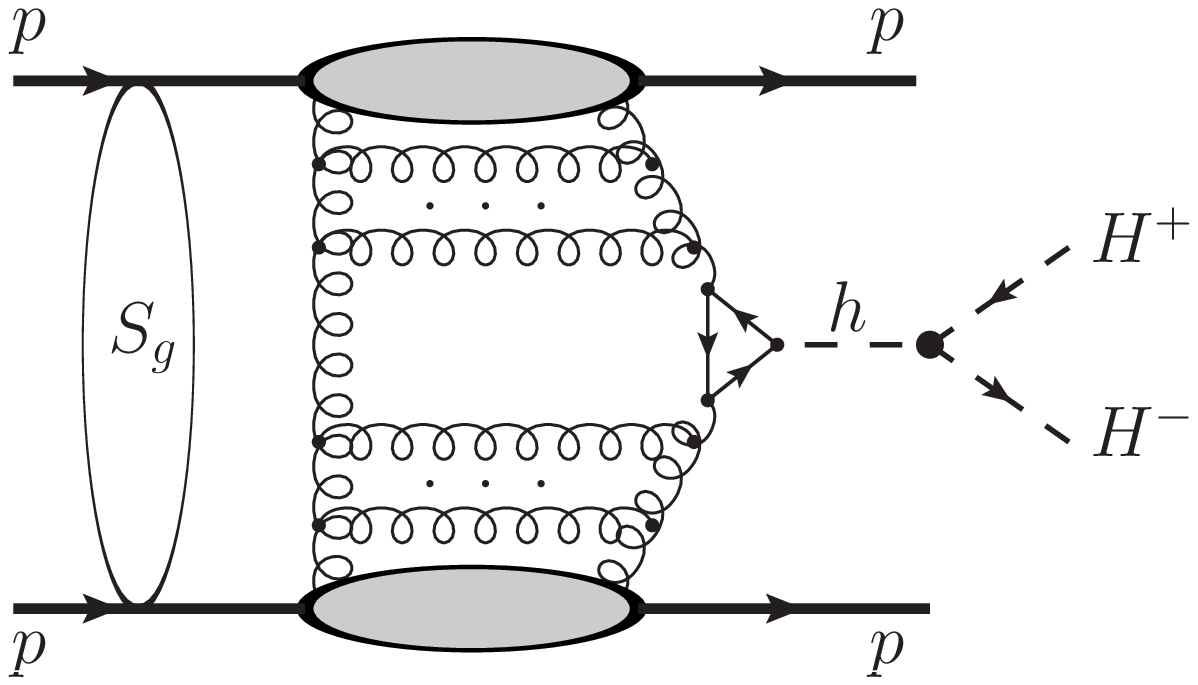}
\caption{\label{fig:diffractive_HpHm}
\small
The diffractive mechanism of the exclusive charged Higgs bosons production
through the intermediate $CP$-even neutral recently discovered Higgs boson.
The absorption corrections due to $pp$ interactions (indicated by the blob)
are relevant at high energies.
}
\end{figure}
So far we have considered a purely electromagnetic process,
the contribution of which is model independent.
The corresponding cross section turned out to be rather low.
Therefore, one could worry whether other processes might not give a sizeable contribution, 
comparable to the photon-photon exchanges. 
One such candidate is the diffractive mechanism discussed, e.g., 
in the context of exclusive Higgs boson production 
\cite{Khoze:1997dr,Khoze:2000cy,Kaidalov:2003ys,Maciula:2010tv}.
In the present case the mechanism shown in
Fig.~\ref{fig:diffractive_HpHm} 
seems an important candidate.
The $g^{*}g^{*} \to H^{+}H^{-}$ hard subprocess amplitude
through the $t$-loop and $s$-channel SM Higgs boson ($h^{0}$) is given by
%
\begin{equation}
V_{g^{*}g^{*} \to H^+ H^-} = 
V_{gg \to h} \, \frac{i}{s_{34} - m_h^2 + i m_{h} \Gamma_{h}} \, g_{h H^+ H^-}
\end{equation}
and enters into ${\cal M}_{pp \to pp H^{+}H^{-}}$
invariant $2 \to 4$ amplitude for the diffractive process
as in \cite{Maciula:2010tv, Lebiedowicz:2012gg}.
The triple-Higgs coupling constant $g_{h H^+ H^-}$ is, of course, model dependent.
In the MSSM model it depends only on the parameters $\alpha$ and $\beta$. 
In the general 2HDM it depends also
on other parameters such as the Higgs potential $\lambda$-parameters 
or masses of Higgs bosons. 
How the coupling constant depends on parameters
of 2HDM was discussed, e.g., in \cite{Gunion:2002zf,Ginzburg:2004vp,Ferrera:2007sp,Baglio:2014nea}.
\begin{figure}[!ht]
\includegraphics[width=0.48\textwidth]{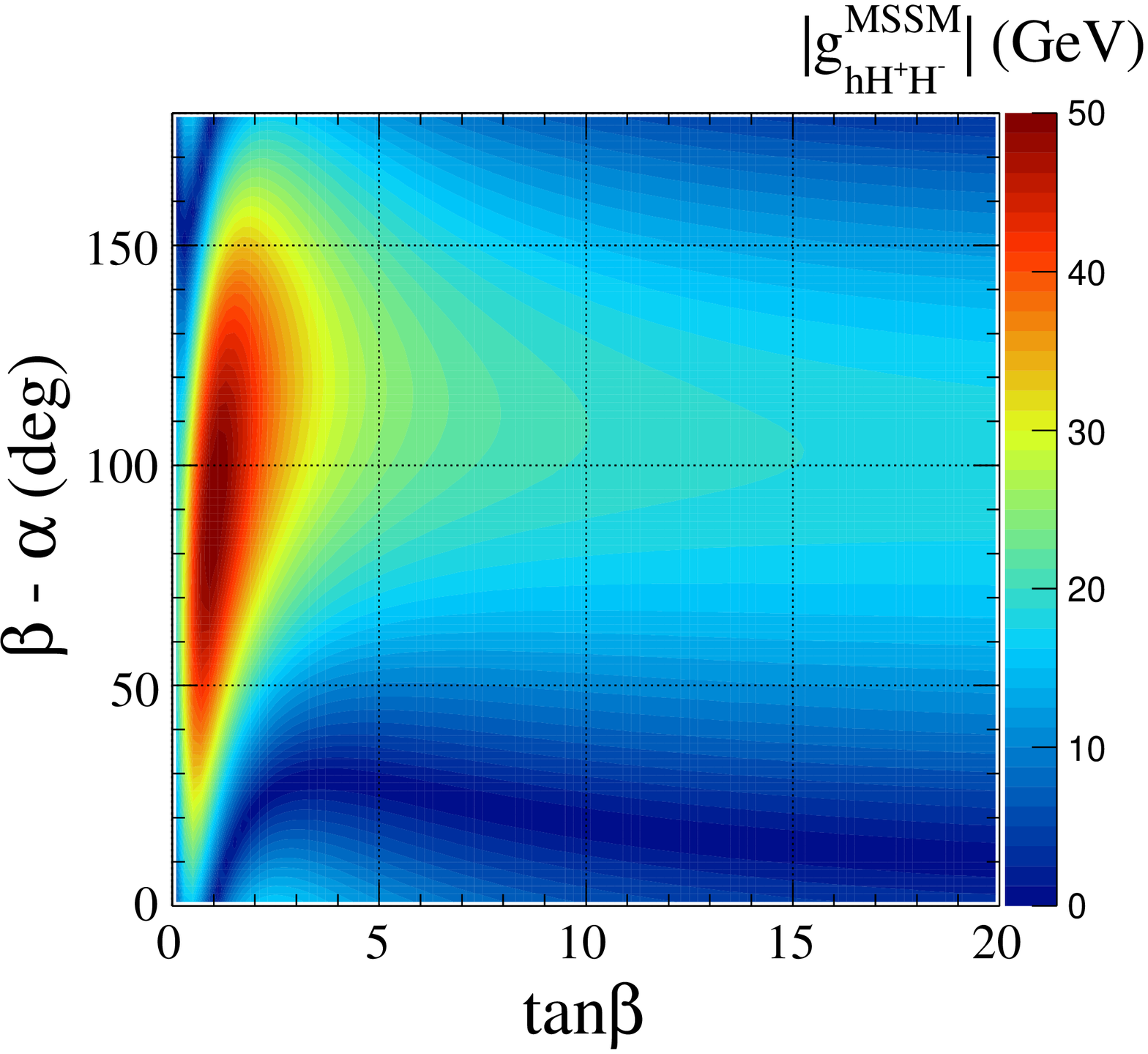}
\includegraphics[width=0.48\textwidth]{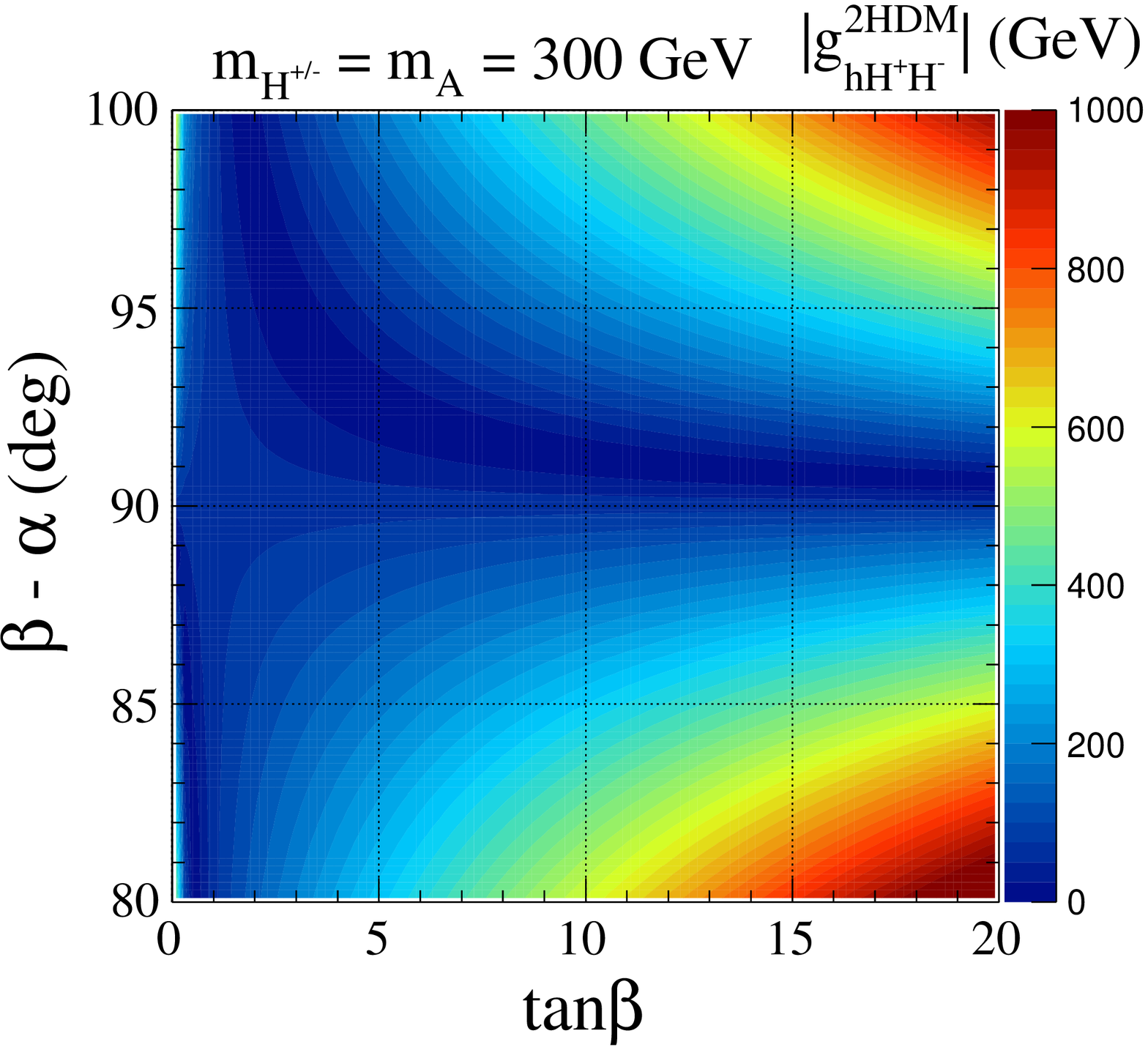}
\caption{\label{fig:g_hHpHm_maps}
\small
The triple-Higgs coupling constant $|g_{hH^+H^-}|$ 
in the [$\tan \beta$, $(\beta - \alpha)$] plane
for the MSSM (left panel) and the type-II 2HDM 
with $m_{H^{\pm}} = m_{A} = 300$~GeV (right panel).
Here we consider the type-II 2HDM 
in which the scalar potential parameters are $Re(\lambda_{5})$ (explicit $Z_{2}$ symmetry) 
and $\lambda_{6} = \lambda_{7} = 0$.
}
\end{figure}
In Fig.\ref{fig:g_hHpHm_maps} we show as an example the coupling
as a function of $\tan \beta$ and $\alpha - \beta$
for MSSM (left panel) and 2HDM (right panel).
In the latter case we have used a relation given in Ref.~\cite{Ferrera:2007sp}
while the formula for the MSSM can be found e.g.~\cite{Gunion:1989we}.
The $g_{hH^+H^-}$ coupling constant in the MSSM case does not exceed 50~GeV
(to be compared e.g. to $g_{hhh} \approx 194$~GeV in the Standard Model).
The coupling constant in the case of 2HDM can be, in general, very large.
Recent data obtained at the LHC in the last three years
put stringent constraints on $\alpha$ and $\beta$ as well as
on masses of the, thus far, unobserved Higgs bosons (some examples
of such analyses can be found in \cite{Coleppa:2013dya,Baglio:2014nea,Broggio:2014mna}.
The LHC experimental data allow for two regions in the [$\tan \beta$, $(\beta - \alpha)$] plane \cite{Eberhardt:2013uba,Baglio:2014nea,Coleppa:2013dya}.
One of them $\beta - \alpha \approx \pi/2$ is the so-called ``alignment limit''.
The second one 
is more difficult to characterize.
In the present analysis we focus on the alignment region
which means the lightest $CP$-even Higgs $h$
is what has been found at the LHC with $m_{h} \simeq 125$~GeV \cite{Aad:2015zhl}.
Experimental data allow for some deviations from the $\beta - \alpha = \pi/2$.
As can be seen from Fig.~\ref{fig:g_hHpHm_maps} a small deviation
from this limit can modify the coupling constant considerably.
The analysis in \cite{Coleppa:2013dya} suggests that $m_{H^{\pm}} \approx m_A$
and we keep such a relation throughout our analysis.
A deviation from such a relation would increase the discussed coupling constant.

\begin{figure}[!ht]
\includegraphics[width=0.48\textwidth]{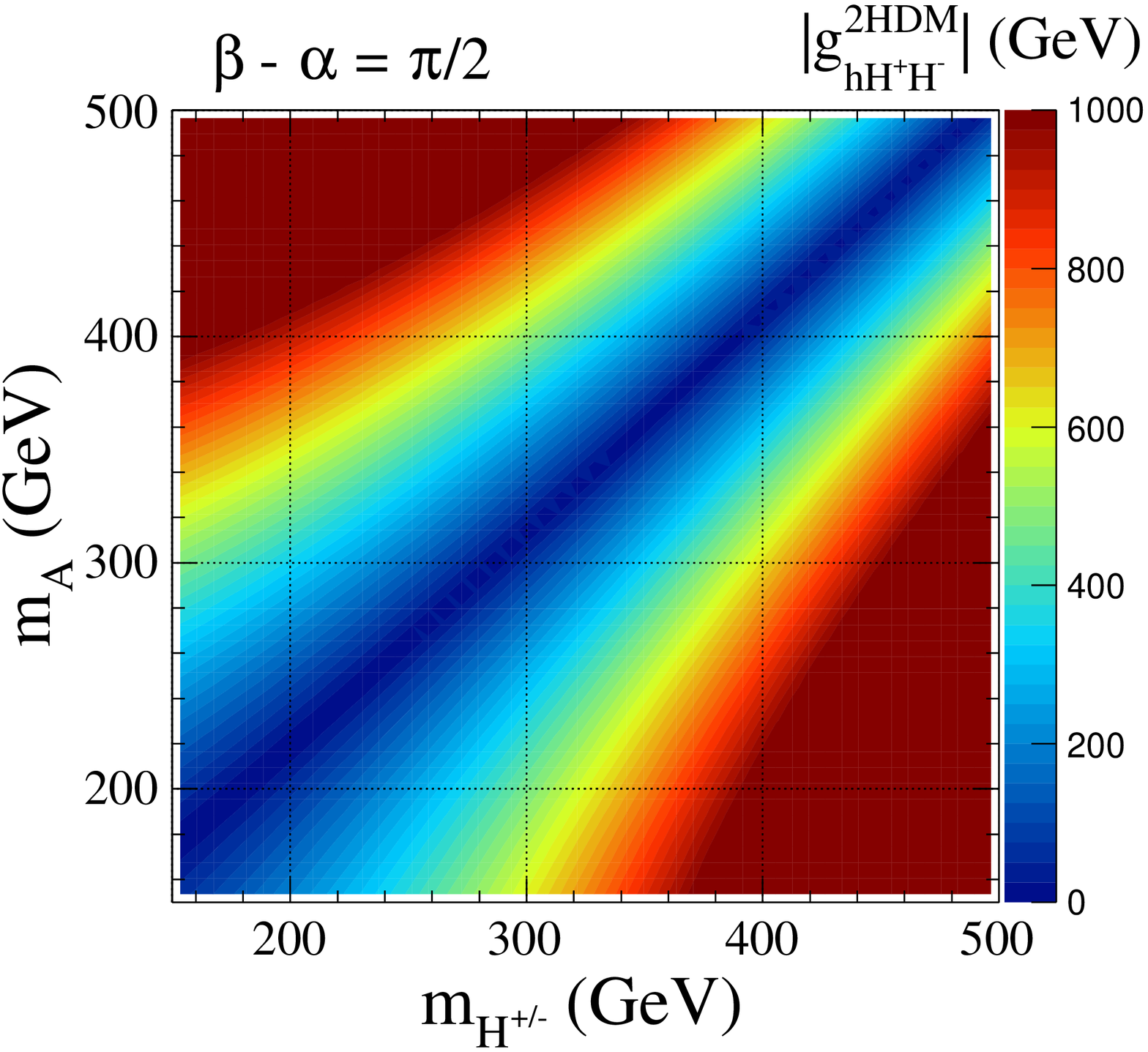}
\caption{\label{fig:MHpmMA}
\small
$|g_{h H^+ H^-}|$ for the 2HDM 
as a function of $m_{H^\pm}$ and $m_A$
for the alignment limit $\beta - \alpha = \pi/2$.
}
\end{figure}
In Fig.~\ref{fig:MHpmMA} we show the dependence of the coupling
constant on masses of charged and $CP$-odd Higgses within 2HDM.
\footnote{We emphasise again that in the MSSM in the lowest order
we have $m_{H^{\pm}}^2 = m_A^2 + m_{W^{\pm}}^2$.}
A minimal value appears when $m_A \approx m_{H^{\pm}}$.
When we relax this condition the coupling can be even as large as 1000~GeV. 
This is consistent with the limits of the allowed region in \cite{Baglio:2014nea}.
Summarizing, $g_{hH^{+}H^{-}}$ in the 2HDM is limited to
64~GeV~$\lesssim g_{hH^{+}H^{-}} \lesssim$~1000~GeV.
The corresponding couplings in the MSSM are smaller than 50~GeV.
\footnote{
It has been shown, e.g., in \cite{Asakawa:2005nx} that in some regions
of the parameter space of 2HDMs the associated production cross section
can be enhanced compared with the MSSM by orders of magnitude.
This is a similar process to that discussed in our paper.}

\begin{figure}
\includegraphics[width=0.48\textwidth]{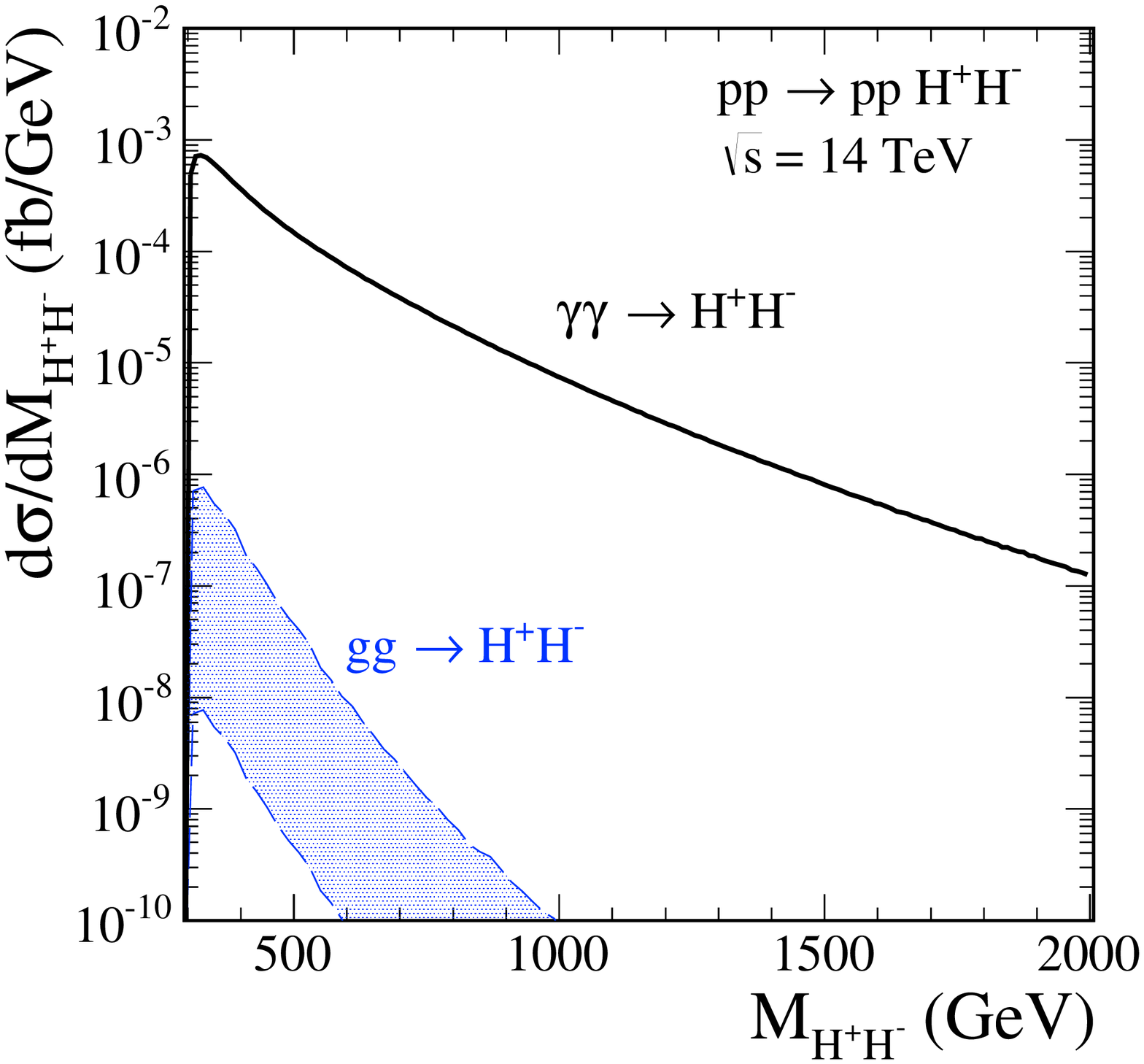}
\includegraphics[width=0.48\textwidth]{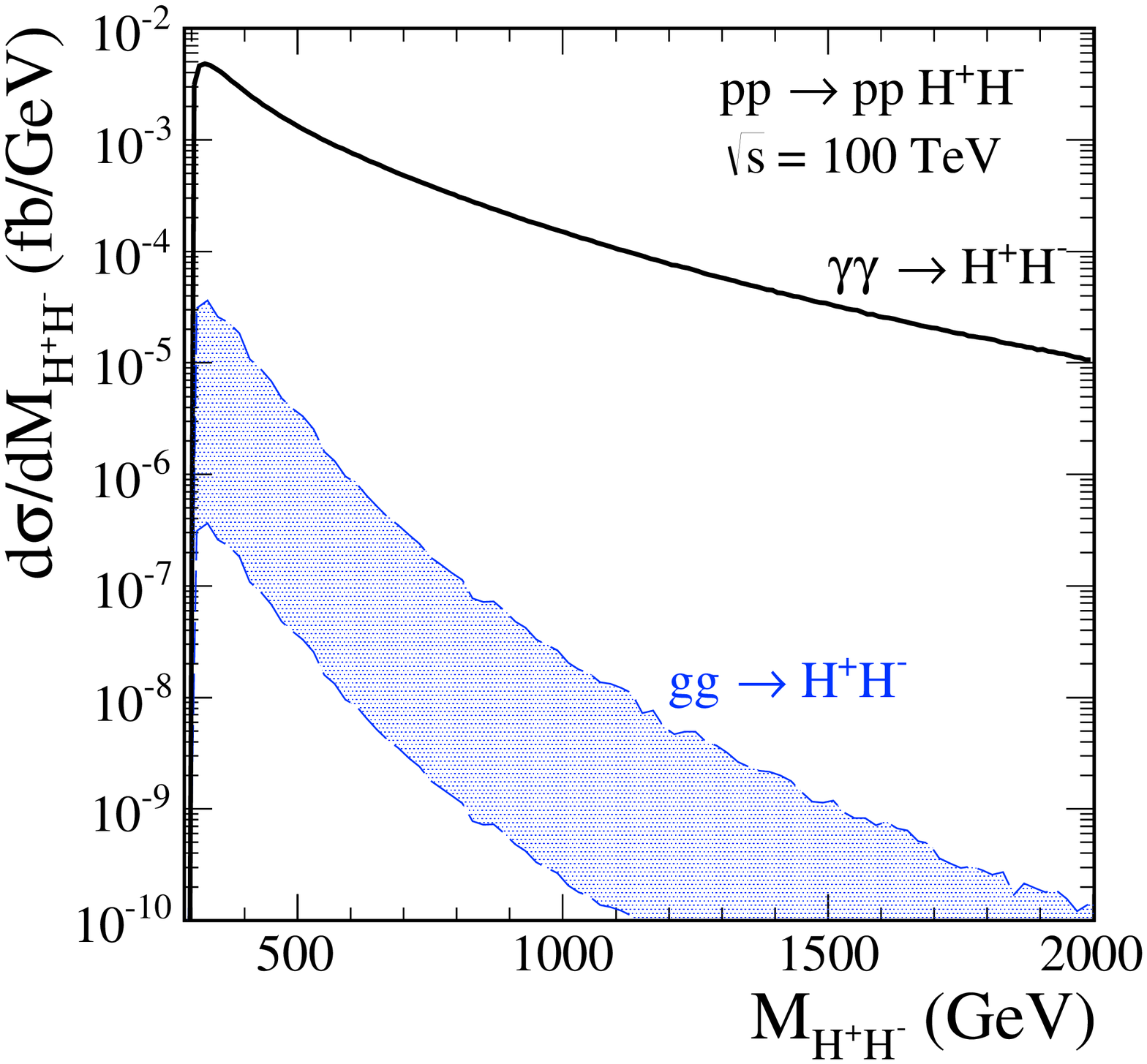}
\caption{
DiHiggs boson invariant mass distributions 
at $\sqrt{s} = 14$~TeV (left panel) and $100$~TeV (right panel).
The upper lines represent the $\gamma \gamma$ contribution.
We also show contribution of the diffractive mechanism (the shaded area)
for the MSTW08 NLO collinear gluon distribution \cite{Martin:2009iq}
and $g_{hH^{+}H^{-}} = 100$ (1000)~GeV for the lower (upper) limit.
}
\label{fig:dsig_dMHH_diffraction}
\end{figure}
In Fig.~\ref{fig:dsig_dMHH_diffraction}
we show corresponding results for the diffractive contribution
for $\sqrt{s}$ = 14 TeV (left panel) and $\sqrt{s}$ = 100 TeV (right panel)
both the lower and upper limit of the 2HDM triple-Higgs coupling
for $m_{H^{\pm}} = 150$~GeV.
In the calculation we have included the ``effective'' gap survival factor 
$\langle S^{2}\rangle=0.03$ typical for the considered range of energies.
The cross section for the exclusive diffractive process
is much smaller than that for $\gamma \gamma$ mechanism
both for LHC and FCC.
In addition to the result for the 2HDM set of parameters
(alignment limit), we also show result with the upper limit $g_{hH^{+}H^{-}} = 1000$~GeV.
With such a big coupling constant, the contribution 
with the intermediate neutral Higgs boson $h^{0}$
dominates over the contribution of boxes for $\tan \beta < 20$. 
Therefore, the upper limit also effectively includes the box contributions
discussed in the context of inclusive $pp \to (gg \to H^{+} H^{-})$ processes \cite{Krause:1997rc,BarrientosBendezu:1999gp,Brein:1999sy}.

\subsection{A comment on possible experimental studies}

So far we have calculated the cross section for the $pp \to pp H^+ H^-$ reaction. 
If one wishes to identify the reaction experimentally, 
one should measure the decay products of $H^{\pm}$ bosons.
The branching fractions to different channels depend on 
the model parameters ($m_{H^{\pm}}$, $\tan \beta$, etc.).
For low masses of $H^{\pm}$ ($m_{H^{\pm}} < 150$~GeV) it is expected that 
the $\tau \nu_{\tau}$ and $c \bar s$ are the dominant channels. 
In the case of heavy charged Higgs (with $m_{H^{\pm}} > 200$~GeV) 
the $t \bar{b} (\bar{t} b)$ or $W^{\pm} h$ channels 
are expected to be the relevant ones. 

In the first case (light $H^{\pm}$) $\tau^+ \tau^-$ could be measured 
in addition to the forward/backward protons.
The emission of neutrinos leads to a strong imbalance
between proton-proton missing mass and $M_{\tau^+ \tau^-}$.
This should help to eliminate the $p p \to p p \tau^+ \tau^-$
reaction, but this requires dedicated Monte Carlo studies,
including perhaps the $pp \to pp \tau^+ \tau^- \gamma$ process.
The $p p \to p p W^+ W^-$ and $p p \to p p H^{\pm} W^{\mp}$
reactions may lead to a similar final state.
Although the branching fraction $W^{+} \to \tau^+ \nu_{\tau}$ or
$W^{-} \to \tau^-  \bar{\nu}_{\tau}$ is only about $\frac{1}{9}$,
it is expected to be a difficult irreducible background because
of the relatively large cross section for the $p p \to p p W^+ W^-$.
In principle, the $c \bar{s} \bar{c} s$ (four jets) final channel 
is also attractive as, in this case, one may check extra conditions 
$M_{q \bar{q}'} - m_{W^{+}}>$~$10-20$~GeV and
$M_{\bar{q} q'} - m_{W^{-}} >$~$10-20$~GeV to exclude 
the $W^+ W^-$ contribution.
The $p p \to p p H^+ W^-$ and $p p \to p p H^- W^+$ processes
may lead to similar final channels ($\tau^{+} \nu_{\tau} \tau^{-} \bar{\nu}_{\tau}$
or $c \bar{s} \bar{c} s$) but the corresponding cross sections
are expected to be smaller (higher-order processes with loops).
Mixed (leptonic $+$ quarkish) final states could also be considered.

In the second case (heavy $H^{\pm}$), in general, both the $t$ quark 
and $b$ jet can be measured. 
In contrast to the previous case we do not know about any sizeable
irreducible background. But then the cross sections are rather
small as discussed in the previous sections.
In the case of the $H^{+} H^{-} \to W^{+} h W^{-} h$ decay channel
the actually measured final state can be rather complicated (e.g., 
$q \bar{q}' b \bar{b} \; q' \bar{q} b \bar{b}$).
Therefore, with experimentally limited geometrical acceptance it may be rather
difficult to reconstruct the charged Higgs bosons.


A detailed analysis of any of the final states considered here 
requires separate Monte Carlo studies including experimental 
geometrical acceptances relevant for a given experiment.
This clearly goes beyond the scope of the present paper which
aims to attract attention to potentially interesting exclusive processes.
The Monte Carlo studies could be done only in close collaboration
with relevant experimental groups.


\section{Conclusions}

In the present paper we have studied in detail the exclusive
production of heavy scalar, weakly interacting, charged bosons in 
proton-proton collisions at the LHC and FCC.
In contrast to EPA our exact treatment of the four-body 
$p p \to p p H^+ H^-$ reaction allows us to calculate any
single particle or correlation distribution.

Results of our exact ($2 \to 4$ kinematics) calculations have been 
compared with those for the equivalent-photon approximation for 
observables accessible in EPA. 
Rather good agreement has been achieved in those cases.
However, we wish to emphasize that some correlation observables
in EPA are not realistic, or even not accessible, 
to mention here only correlations in azimuthal angle 
between the outgoing protons or the charged Higgs bosons.
We have predicted an interesting minimum at $\phi_{pp}=90^o$
which is a consequence of the field theoretical couplings
involved in the considered reaction.

We have analyzed in detail the role of the Dirac and Pauli
form factors. In contrast to light particle production,
the Pauli form factor plays an important role especially 
at large $M_{H^{+}H^{-}}$,
and related terms in the amplitude cannot be neglected.
We see that the double spin preserving
contributions are almost identical in both exact and EPA calculations
(within $1 \%$), but the spin-flipping contributions are in 
our calculation somewhat smaller. 

In the present paper we have studied, for the first time for the considered
two-photon-induced reaction, the absorption effects due to proton-proton
(both initial and final state) nonperturbative interactions.
Any extra interaction may, at the high energies, lead to a production
of extra particles destroying exclusivity of the considered reaction.
The absorptive effects lead to a reduction of the cross section.
The reduction depends on kinematical variables.
A good example are distributions in four-momentum transfers squared.
At small $|t_1|$ and $|t_2|$, the absorption is weak and increases 
when they grow. 
We have also found interesting dependence of the absorption on $M_{H^{+}H^{-}}$. 

The relative effect of absorption is growing with growing $M_{H^{+}H^{-}}$.
A similar tendency has been predicted recently for the 
$p p \to p p W^+ W^-$ in the impact parameter approach \cite{Dyndal:2014yea}. 
The impact parameter approach is, however, not useful for many observables studied here.
We have predicted that the absorption effects for our two-photon-induced
process become weaker at larger collision energy which is in contrast
to the typical situation for diffractive exclusive processes.
Our study shows that an assumption of no absorption or constant (small)
absorption effects, often assumed in the literature for photon-photon-induced processes, 
is rather incorrect and corresponding results should be corrected.

In addition to calculating differential distributions corresponding to 
the $\gamma \gamma$ mechanism we have performed first calculations
of the $H^+ H^-$ invariant mass for the diffractive KMR mechanism. 
We have tried to estimate limits on 
the $g_{h H^+ H^-}$ coupling constant within 2HDM based on recent
analyses related to the Higgs boson discovery.
The diffractive contribution, even with the overestimated
$|g_{h H^+ H^-}|$ coupling constant, gives a much smaller cross section
than the $\gamma \gamma$ mechanism.
We have also made an estimate of the contributions related
to $\gamma Z$, $Z \gamma$, and $Z Z$ exchanges and found that their
contributions are completely negligible.
This shows that the inclusion of the $\gamma \gamma$ mechanism should
be sufficient, and the corresponding cross sections should be reliable.

Whether the $p p \to p p H^+ H^-$ reaction can be identified at 
the LHC (run 2) or FCC requires further studies including simulations 
of the $H^{\pm}$ decays. 
Two $H^{\pm}$ decay channels seem to be worth studying
in the case of light $H^{\pm}$:
$H^{\pm} \to \tau^{+} \nu_{\tau} (\tau^{-} \bar{\nu}_{\tau})$ or $H^{\pm} \to c \bar s (\bar c s)$.
The first decay channel may be difficult due to a competition
of the $p p \to p p W^+ W^-$ reaction which can also contribute to 
the $\tau^+ \tau^-$ channels. 
The combined branching fraction is about 0.11$^2$ = 0.0121 
(two independent decays) which is not so small given the fact 
that the cross section for the $W^+ W^-$ production 
is much bigger than that for $H^+ H^-$ production.
In the second case (four quark jets), one could measure
invariant masses of all dijet systems to reduce the $W^+ W^-$ background.
In the case of the heavy $H^{\pm}$ Higgs boson,
the $H^{\pm} \to t \bar{b} (\bar{t} b)$ decay can be considered.
In principle, both the $t$ quark and $b$ jet can be measured. 
In this case we do not know about any sizeable irreducible background.

The reaction considered in this paper is a prototype for any two-photon-induced process. 
In the future we wish to also consider the $p p \to p p W^+ W^-$ reaction 
where similar effects may occur.
This reaction was proposed to search for the anomalous triple or quartic 
boson coupling.
Effects beyond the Standard Model are expected at rather large 
invariant masses $M_{W^+ W^-}$, where we have found
strong absorptive corrections. This exclusive reaction is, however, 
more complicated due to the more complex couplings and spins involved 
and due to weak decays of the two $W$ bosons 
where strong spin-spin correlation effects are expected. 

\acknowledgments
We are indebted to Jan Kalinowski for a discussion and particularly 
to Maria Krawczyk for help in understanding the present limitations
of the 2HDM.
The help of Rafa{\l} Maciu{\l}a in calculating the diffractive component
is acknowledged.
The work of P.L. was supported by the Polish NCN Grant No. DEC-2013/08/T/ST2/00165 (ETIUDA) 
and by the MNiSW Grant No. IP2014~025173 ``Iuventus Plus''
as well as by the START fellowship from the Foundation for Polish Science.
The work of A.S. was partially supported by the Polish NCN Grant No.
DEC-2011/01/B/ST2/04535 (OPUS) as well as by the Centre for Innovation and
Transfer of Natural Sciences and Engineering Knowledge in Rzesz\'ow.

\include{bibliography}

\end{document}

%% file: bibliography.tex
\nocite{}

{
\begin{small}
\bibliography{refs}
\end{small}
}